\documentclass[sigconf,10pt,noacm]{acmart}
\usepackage{acmart-tight-wqz0226}



\usepackage{tikz}

\usepackage{booktabs} 
\usepackage{import}
\usepackage{mathtools}
\usepackage{subfigure}
\graphicspath{{Fig/}}
\usepackage{booktabs} 
\usepackage{algorithm}
\usepackage{algorithmicx} 
\usepackage{algpseudocode}
\usepackage{paralist}
\usepackage{subcaption}
\usepackage[most]{tcolorbox}
\tcbuselibrary{skins, breakable}
\usepackage{multirow}
\usepackage{xspace}
\usepackage{subfigure}
\usepackage{soul}
\usepackage{array}
\usepackage{diagbox}
\usepackage{verbatim} 
\usepackage{colortbl}
\usepackage{upgreek}
\usepackage{enumitem}
\usepackage{svg}
\usepackage{gensymb} 
\usepackage{bm}
\usepackage{subcaption}
\usepackage{adjustbox}
\usepackage{geometry}
\usepackage{tabularx}
\usepackage{amsthm}

\usepackage{url}
\usepackage{twemojis}

\definecolor{ao}{rgb}{0.54, 0.17, 0.89}

\newcommand{\name}{\textsc{RFDT}\xspace}

\def\ie{\textit{i.e.}\xspace}
\def\eg{\textit{e.g.}\xspace}
\def\etc{\textit{etc.}\xspace}

\def\wrt{\textit{w.r.t.}\xspace}
\newcommand{\ttf}[1]{\textbf{\textit{#1}}}

\definecolor{linkblue}{RGB}{20,60,180}
\newcommand{\revise}[1]{{\color{black}{#1}}}
\definecolor{Melon}{RGB}{208, 76, 23}
\newcommand{\revisetwo}[1]{{\color{black}{#1}}}
\definecolor{Orchild}{RGB}{180, 84, 187}
\newcommand{\revisethree}[1]{{\color{black}{#1}}}

\newtcolorbox{highlightbox}[1][]{
  enhanced,
  breakable,
  colback=gray!10,
  colframe=gray!50,
  coltitle=black,
  fonttitle=\bfseries,
  title=#1,
  boxrule=1pt,
  arc=2mm,
  outer arc=2mm,
  top=6pt,
  bottom=6pt,
  left=6pt,
  right=6pt
}

\tcbset{
  myformula/.style={
    enhanced,
    colframe=orange!50!black,
    colback=white,
    boxrule=0.4pt,
    arc=2mm,
    top=4pt,
    bottom=4pt,
    left=6pt,
    right=6pt,
    boxsep=2pt,
    sharp corners=south,
    attach boxed title to top center={yshift=-\tcboxedtitleheight/2},
    boxed title style={
      colframe=white,
      colback=white,
      font=\small\itshape,
    }
  }
}
\begin{document}



\title[RF Digital Twin]{Physically Accurate Differentiable Inverse Rendering\texorpdfstring{\\}{ }for Radio Frequency Digital Twin}%


\author{\texorpdfstring{Xingyu~Chen$^{\dagger}$, Xinyu~Zhang$^{\dagger}$, Kai~Zheng$^{\dagger}$, Xinmin Fang$^{\ddagger}$,  Tzu-Mao~Li$^{\dagger}$,~Chris~Xiaoxuan Lu$^{\S}$, Zhengxiong~Li$^{\ddagger}$}{Xingyu Chen, et al.}} 
\affiliation{%
  \institution{$^{\dagger}$University of California, San Diego, $^{\ddagger}$University of Colorado Denver, $^{\S}$University College London}%
\country{}
}
\renewcommand{\shortauthors}{Xingyu Chen, et al.\@}

\settopmatter{printacmref=false, printccs=false, printfolios=true}
\renewcommand\footnotetextcopyrightpermission[1]{}




\begin{abstract}


Digital twins, virtual simulated replicas of physical scenes, are transforming system design across industries. However, their potential in radio frequency (RF) systems has been limited by the non-differentiable nature of conventional RF simulators. The visibility of propagation paths causes severe discontinuities, and differentiable rendering techniques from computer graphics are no longer applicable due to point-source antennas and dominant specular reflections. In this paper, we present \name, a physically based differentiable RF simulation framework that enables gradient-based interaction between virtual and physical worlds. \name resolves discontinuities with a physically grounded edge-diffraction transition function, and mitigates non-convexity from Fourier-domain processing through a signal-domain transform surrogate. Our implementation demonstrates \name's ability to accurately reconstruct digital twins from real RF measurements. 
Moreover, \name can augment diverse downstream applications, such as test-time adaptation of machine learning-based RF sensing and physically constrained optimization of communication systems. 
The supplementary materials and code of \name are released at 
\textcolor{magenta}{\href{https://RFDigitalTwin.github.io/}{rfdigitaltwin.github.io}}.




\end{abstract}







\maketitle

\section{Introduction}

Radio Frequency (RF) signals serve not only as a medium for communication but also interact richly
with the physical world. Subtle signal propagation and
reflection patterns can encode information about the environment.  This
property has powered a wave of innovative RF sensing applications, including
micro-gesture detection \cite{zheng2019zero,lien2016soli,li2022towards,wei2015mtrack}, vital
sign monitoring \cite{xu:mobicom2017,hwang2022solving}, and non-line-of-sight (NLOS) imaging
\cite{zhao2018rf, zhao2019through, xue2021mmmesh, woodford2023metasight}. Likewise, wireless
communication systems leverage multipath propagation to enhance reliability,
coverage, and spatial multiplexing gain. This is particularly critical in the millimeter-wave (mmWave) bands where performance becomes highly
sensitive to ambient environment \cite{EMi}. 

To model and leverage such RF-environment interactions, conventional
system design has largely followed two paths.  Data-driven approaches can
extract complex patterns from large datasets, but often fail to
generalize to unseen environments \cite{guan2020through,xue2021mmmesh,Wang2022WiMesh,chen2023rfgenesis}. Model-driven approaches \cite{chen2023environment, qian20203d}, rooted
in RF propagation theory and signal processing, provide stronger generalization  
but cannot incorporate the implicit, experiential knowledge about
physical environments. 

\begin{figure}[t]
\centering
\includegraphics[width=0.48\textwidth]{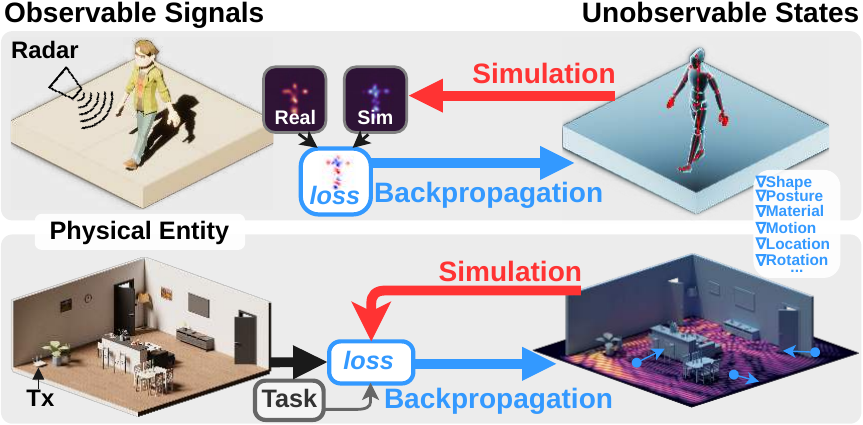}

\caption{\name constructs digital twins by solving the inverse problem of RF simulation. Top: 3D reconstruction through radar simulation. Bottom: Communication system optimization through RF scene rendering.}
\label{fig:pull}
\end{figure} 

To propel RF system design to its next evolutionary phase, in this paper, we propose a novel framework that casts the problem as constructing digital twins (DTs) of objects in the physical world. 
As illustrated in Fig.~\ref{fig:pull}, a DT achieves high-fidelity simulation through continuous, bidirectional, 
and gradient-based interactions with its physical counterpart. This interaction distinguishes DTs from unidirectional methods such as conventional RF ray tracing simulation (digital-to-physical) or sensing systems (physical-to-digital).
Our DT framework, referred to as \textit{\textbf{\name}}, realizes such interaction by incorporating 
both into a closed-loop optimization framework. 
Unlike conventional forward simulation
techniques \cite{chen2023rfgenesis} that use ray tracing to predict the RF signal patterns (output)
from the physical twin's characteristic parameters (input), \name solves the \textit{inverse problem}, \ie, inferring the unknown input from the observable output. 
Since the DT model is physics grounded and explainable, it overcomes the
reliance on training data or labels, and intrinsically handles \textit{unseen} 
scenes. On the other hand, its gradient-based optimization enables easy
integration of implicit knowledge and constraints (\eg, possible geometries
for given object categories) represented by pre-trained neural blocks.

To enable the closed-loop optimization in DT, the key challenge lies in 
\textit{\textbf{differentiability}} of simulation steps that characterize how the RF signals propagate and interact with physical objects. 
The output of a differentiable RF simulation should vary smoothly and predictably in response to infinitesimal perturbations of its input parameters (\eg, the geometry or orientation of a reflecting surface). This property enables accurate computation of gradients between the simulation output and the input parameters, which is fundamental for efficient and robust DT optimization. 

While differentiable rendering in computer graphics has achieved remarkable progress, we prove in Sec.~\ref{sec:discontinuity} that these methods cannot be directly applied to RF. 
Graphics rendering heavily relies on Monte Carlo integration, which estimates scene illumination by averaging many randomly chosen light paths.  
In typical optical settings, light sources are modeled as extended area emitters, and objects' surfaces scatter rays over many directions. In RF, by contrast, antennas behave as near-point sources, and ray-surface interaction is dominated by specular reflections (Sec.~\ref {sec:discontinuity}).  As a result, the random path sampling in Monte Carlo integration is almost surely to miss the dominant specular path. 
In addition, many direct differentiation schemes for geometric parameters in ray
tracing rely on aggressive approximations \cite{liu2019soft}, resulting in
inaccurate gradient estimates. This is especially problematic in RF
optimization, where wavelength-scale precision is essential.

Our \name framework represents the first to establish physically consistent gradients for RF simulation. It enables fully differentiable RF ray tracing by identifying and solving two fundamental challenges.

\textit{\ttf{(i)} Discontinuity with respect to scene parameters.} 
In traditional RF ray tracing, a tiny perturbation in geometry or material parameters can abruptly change which propagation paths contribute to the received signal, resulting in sharp discontinuities in both amplitude and phase.
To overcome this challenge, 
we reformulate the RF ray tracing process into a continuous function. 
In particular, instead of using the conventional binary path visibility test that either fully includes or fully discards a reflection path, we introduce a physically grounded transition function based on the principle of equivalent edge diffraction.  
This models how path contributions vary smoothly as reflection points approach or cross object boundaries, making the dominant specular effects differentiable and ensuring stable gradient propagation throughout the simulation.

\textit{\ttf{(ii)} Local non-convexity caused by the RF waves' periodic properties and signal-domain transforms.} The inherent periodicity of RF waves introduces non-convexity into simulations of signal properties such as strength and angle. This issue is compounded by the signal-domain transforms commonly used in radar and communication systems (\eg, FFT for multi-carrier modulation and range/frequency analysis), which further create spurious sidelobes and local minima. As a result, direct application of gradient descent on RF signals, particularly radar waveforms widely used in RF sensing, often fails to converge because the optimization landscape is filled with wavelength-scale ambiguities. To address this challenge, we incorporate a novel surrogate model for transform-domain signal representations, in particular the radar range profile. This surrogate not only provides smooth and accurate gradients but also significantly reduces the computational overhead of the \name optimization framework.

\begin{table*}[t!]
\footnotesize
  \caption{Comparison of \name{} with related works in Differentiable Rendering}
  \label{tab:related_comparison}
  \addtolength{\tabcolsep}{-.2pt}%
  \begin{tabularx}{\linewidth}{  
  c|
  cc|
  ccc|
  cc|
  >{\centering\arraybackslash}p{0.069\linewidth}c}
    \toprule
    \textbf{Work}
      & \textbf{Spectrum}
      & \textbf{RT Method}
      & {\scriptsize\textbf{Point Source}}
      & {\scriptsize\textbf{Spec.Reflec.}}
      & {\scriptsize\textbf{Diffraction}}
      & \textbf{$\partial$-Geometry}
      & \textbf{Unbiased $\nabla$}
      & \textbf{Radar}
      & \textbf{Comm.} \\
    \midrule
        Direct AD~\cite{NimierDavidVicini2019Mitsuba2}
      & Optical
      & Monte Carlo
      & \twemoji{cross mark} No
      & \twemoji{cross mark} No
      & \twemoji{cross mark} No
      & \twemoji{cross mark} No
      & \twemoji{cross mark} No
      & \twemoji{cross mark} No
      & \twemoji{cross mark} No \\
    Edge Sampling~\cite{li2018differentiable}
      & Optical
      & Monte Carlo
      & \twemoji{cross mark} No
      & \twemoji{cross mark} No
      & \twemoji{cross mark} No
      & \includegraphics[width=0.9em]{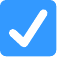} Yes
      & \includegraphics[width=0.9em]{Figures/blue_check.pdf} Yes
      & \includegraphics[width=0.9em]{Figures/blue_check.pdf} Yes
      & {\scriptsize\twemoji{small orange diamond} Potential}
      \\
    Path Space~\cite{Zhang:2020:PSDR,zhang2021path}
      & Optical
      & Monte Carlo
      & \includegraphics[width=0.9em]{Figures/blue_check.pdf} Yes
      & \twemoji{cross mark} No
      & \twemoji{cross mark} No
      & \includegraphics[width=0.9em]{Figures/blue_check.pdf} Yes
      & \includegraphics[width=0.9em]{Figures/blue_check.pdf} Yes
      & {\scriptsize\twemoji{small orange diamond} Potential}
      & {\scriptsize\twemoji{small orange diamond} Potential} \\
    Warped Area~\cite{bangaru2020unbiased,loubet2019reparameterizing}
      & Optical
      & Monte Carlo
      & \twemoji{cross mark} No
      & \includegraphics[width=0.9em]{Figures/blue_check.pdf} Yes
      & \twemoji{cross mark} No
      & \includegraphics[width=0.9em]{Figures/blue_check.pdf} Yes
      & \includegraphics[width=0.9em]{Figures/blue_check.pdf} Yes
      & {\scriptsize\twemoji{small orange diamond} Potential}
      & {\scriptsize\twemoji{small orange diamond} Potential} \\
      \midrule
    MetaWave~\cite{chen2023metawave}
      & Radio Freq.
      & Monte Carlo
      & \includegraphics[width=0.9em]{Figures/blue_check.pdf} Yes
      & \twemoji{cross mark} No
      & \twemoji{cross mark} No
      & \includegraphics[width=0.9em]{Figures/blue_check.pdf} Yes
      & \twemoji{cross mark} No
      & \includegraphics[width=0.9em]{Figures/blue_check.pdf} Yes
      & \twemoji{cross mark} No \\
    Sionna~\cite{sionna}
      & Radio Freq.
      & \revise{Both}
      & \includegraphics[width=0.9em]{Figures/blue_check.pdf} Yes
      & \includegraphics[width=0.9em]{Figures/blue_check.pdf} Yes
      & \includegraphics[width=0.9em]{Figures/blue_check.pdf} Yes
      & \twemoji{cross mark} No
      & \twemoji{cross mark} No
      & \twemoji{cross mark} No
      & \includegraphics[width=0.9em]{Figures/blue_check.pdf} Yes \\
    \rowcolor{gray!10}
    \textbf{\name} (Ours)
      & Radio Freq.
      & Both
      & \includegraphics[width=0.9em]{Figures/blue_check.pdf} Yes
      & \includegraphics[width=0.9em]{Figures/blue_check.pdf} Yes
      & \includegraphics[width=0.9em]{Figures/blue_check.pdf} Yes
      & \includegraphics[width=0.9em]{Figures/blue_check.pdf} Yes
      & \includegraphics[width=0.9em]{Figures/blue_check.pdf} Yes
      & \includegraphics[width=0.9em]{Figures/blue_check.pdf} Yes
      & \includegraphics[width=0.9em]{Figures/blue_check.pdf} Yes \\
    \bottomrule
  \end{tabularx}%
    \begin{center}\footnotesize
    \textbf{$\partial$-Geometry: Differentiable \wrt Geometry;\quad
    Spec.Reflec.: Specular Reflection; \quad
    Comm.: Communication;\quad
    Radio Freq.: Radio Frequency.}
  \end{center}
\end{table*}

With the power of differentiability, our \name framework implements an iterative algorithm to optimize the DT's parameters, guided by the backpropagated gradients (Fig.~\ref{fig:pull}).
Unlike data-driven approaches, \name \textit{explicitly} parameterizes the
physical scene with visual attributes (\eg, geometry, position, pose) and RF
attributes (\eg, material properties, radar configuration), ensuring
interpretability and generalization to unseen settings. Unlike purely
model-driven approaches, \name can embed pre-trained neural networks as
\textit{implicit knowledge} to regularize the optimization with learned priors
of physical scenes. 

We implement \name as a differentiable RF ray tracing
engine, built from scratch in optimized C++/CUDA with native PyTorch support. THe system includes  
custom gradient computation modules for path differentiation and surrogate signal
transforms. Our evaluation shows that \name produces accurate gradients that
closely match the ground truth. Using real radar measurements, \name
reconstructs 3D shapes (vehicles, human postures, and daily objects) with high
accuracy. It supports multi-object and NLOS scenes, and can leverage prior
knowledge of object categories to further enhance fidelity and accelerate
convergence. In addition, we present case studies to demonstrate \name's
versatility: test-time adaptation for machine learning-based RF sensing systems,
and physically constrained optimization of RF communication parameters.
In both cases, \name is shown to enhance the performance of state-of-the-art
data-driven and model-driven systems.

The key contributions of \name are as follows:

\noindent $\bullet$ 
We propose \name, a novel simulation-in-the-loop optimization framework that enables bidirectional, physics-grounded interaction between DTs and corresponding physical twins. 

\noindent $\bullet$ 
We design a physically grounded differentiable RF ray tracing framework that overcomes discontinuity and local non-convexity challenges, enabling RF simulation to be integrated within end-to-end system optimization.

\noindent $\bullet$ 
We provide a complete implementation of \name. Our extensive evaluation and case studies demonstrate how \name accurately reconstructs DTs, while augmenting existing data-driven and model-driven systems.

\section{Background and Related Work}
\label{sec:background}

\textbf{RF Simulation Pipeline.}
RF propagation simulation has been extensively studied, with ray tracing as a fundamental technique.
Ray tracing uses \textit{scene state} as input, represented by a set of \textit{scene parameters} $\theta$, characterizing the geometry and material properties of the RF scene. 
In practice, the scene is often discretized into tiny \textit{mesh} elements
(\eg, triangles) to facilitate computational analysis. 
The core process of ray tracing is to compute propagation paths between transmitter and receiver that are \textit{physically valid}, in the sense that they obey the geometric laws of propagation, including reflection, refraction, and diffraction.

For specular reflections, ray tracing commonly uses the classical method-of-images \cite{thomson1845mathematical,balanis2012advanced},
which models reflections by introducing virtual mirrored sources, transforming each reflection into a straight-line path in an equivalent virtual scene.
Material interactions and NLOS propagation are modeled
using Fresnel coefficients, which capture amplitude attenuation and phase shift
at dielectric boundaries. Refracted paths follow Snell's law as rays enter and
exit objects. 
Our \name simulator also supports heterogeneous internal propagation to capture
multilayer structures, analogous to volumetric rendering in graphics where rays
accumulate effects while traversing a medium (details in
Appendix~\ref{appendix:material_nlos}).

For each valid path $i$, the RF simulation outputs its channel impulse response
(\ttf{CIR}), characterized by delay $\tau_i$, amplitude $\alpha_i$, and phase
shift $\phi_i$. The received CIR is the superposition of all the multipath
contributions, which can be further transformed into structured representations
such as \textit{spatial spectrum}, \ie, distribution of signal strength in 3D
space. Note that, unlike full-wave electromagnetic simulation based on Maxwell's
equations \cite{maxwell1865viii}, RF ray tracing generally cannot reproduce the
absolute phase of received signals, as it ignores fine-grained object-wave
interactions (\eg, near-field coupling) and RF hardware
artifacts. However, it can capture the relative phase between paths,
which indirectly affect the signal strength components of the CIR through
constructive and destructive multipath interference.

\textbf{Differentiable rendering techniques in computer graphics (CG)}
can retrieve the physical parameters of objects (\eg, 3D geometries) in a scene from a sparse set of images \cite{marschner1998inverse}. 
Early research relied on partial knowledge of the scene, \eg, lighting \cite{marschner1997inverse}, texture \cite{wei2008inverse}, and material \cite{chen2020invertible}. 
Neural rendering techniques can implicitly represent 3D scenes and synthesize novel views through volumetric ray marching \cite{mildenhall2020nerf}. However, these methods lack an explicit 3D representation.
More recently developed optical differentiable renderers  \cite{kato2020differentiable, petersen2022gendr, loubet2019reparameterizing}, 
summarized in \ttf{Table~\ref{tab:related_comparison}}, 
extend the classical CG pipeline to support gradient-based optimization over explicit 3D structures such as polygon meshes.

These CG techniques typically rely on Monte Carlo integration to estimate how
light propagates in a scene. They are effective because a light source
is often modeled as extended emitters that shoot rays from many origin points on
the source's body, and each rays' interaction with a surface tends to scatter
light into all directions. Since enumerating all possible light
paths is infeasible, Monte Carlo sampling provides a practical solution: it
draws a large number of trial paths at random and averages their contributions,
yielding an unbiased estimator of the underlying \textit{rendering equation}
\cite{kajiya1986rendering}---the integral describing how light propagates
and reflects in space. However, this method is fundamentally ineffective for specular reflection and point sources (Sec.~\ref{sec:discontinuity}).

\textbf{Inverse and differentiable RF simulation.}
Research on inverse and differentiable RF simulation is still at an early stage. Existing approaches either rely on biased approximations, or lack the ability to differentiate with respect to geometric scene parameters, which makes them fundamentally unsuitable for closed-loop RF system optimization.

MetaWave \cite{chen2023metawave} represents an early attempt at using differentiable simulation to design reflecting objects for radar deception. However, its simulation simplifies the RF wave properties, focusing solely on the signal strength of first-order reflections. 
Nvidia's Sionna \cite{sionna} represents the state-of-the-art in RF ray tracing. 
However, Sionna is mainly designed for modeling the communication channel distributions, and cannot deal with the differentiability challenges arising from RF interactions with geometric scene parameters.   
Our experiments (Sec.~\ref{sec:convergeEval}) show that Sionna's gradient computation \wrt geometry breaks down for RF sensing tasks.  

\textbf{RF Sensing for object characterization and reconstruction.}
Substantial recent research has explored RF sensing for reconstructing 3D
objects and characterizing their geometries and materials.  The resolution of
traditional signal processing methods is fundamentally limited by the radar
antenna aperture and sampling bandwidth \cite{qian20203d}. 
Data-driven methods overcome the limitation by incorporating pre-trained
knowledge of object characteristics
\cite{zhao2019through,guan2020through,xue2021mmmesh,ren20213d,jiang2020towards,Wang2022WiMesh,kong2022m3track,Xue2022M4esh}.
For example, they can reconstruct human postures \cite{xue2021mmmesh} and
vehicles \cite{guan2019high} by fitting sparse radar point cloud to learned
meshes.  These approaches depend heavily on training data.  Generative models
can synthesize diverse training data \cite{chen2023rfgenesis} to improve
generalization. However, these models themselves need training, and they cannot
easily adapt to site-specific multipath environment or uncommon object
properties (Sec.~\ref{sec:usecase1}).  In contrast, \name is a DT optimization framework that can fit the
physical scene without training or fine-tuning. 


\vspace{-2mm}
\section{\name Differentiable Rendering}
\label{sec:diffRT}

\subsection{Discontinuity Challenges}
\label{sec:discontinuity}

Extending differentiable rendering into the RF domain can be viewed as making
traditional RF ray tracing differentiable.  
This extension faces two unique challenges rooted in RF propagation physics. 

\textit{First}, RF antennas are typically sub-wavelength in size, acting
effectively as point sources. Also, the much longer RF wavelength makes
surface interactions predominantly specular. Therefore, when RF source
projects signals towards a surface, the valid reflection paths are confined to a
\textit{measure-zero} set of directions.
So random sampling methods (\eg, Monte Carlo integration) are almost
certain to miss them.
This measure-zero issue is inherent to all ray tracers with point sources and
specular reflections. 
What sets \name apart lies in its need to produce accurate and stable gradients for gradient-based optimization, which conventional RF ray tracers are not designed to provide.

\textit{Second}, since specular reflection redirects a ray into only one
direction upon surface interaction, even tiny geometric perturbations can
abruptly switch a path from valid to invalid. The path \textit{visibility
function} 
indicates whether a candidate Tx-surface-Rx path is valid and unobstructed. 
In ray-tracing, this function is evaluated via a sequence of \textit{triangle
hit tests}: if every intersection along a path is valid, the function returns
1; and 0 otherwise. As a result, the visibility function behaves as a step
function of scene geometry, producing \ttf{discontinuities} and ill-defined
gradients. In what follows, we prove that differentiable renderers based on Monte Carlo integration are fundamentally incompatible with RF. 

\vspace{-3mm}
\begin{highlightbox}
\small
\textbf{Proposition:} Naïve Monte Carlo integration fails to resolve the radiance from a perfect specular surface illuminated by a point source. Because the interaction is characterized by \textbf{Dirac delta} distributions, the valid reflection paths constitute a set of \textbf{measure zero} within the sampling domain, resulting in a zero probability of capturing any energy (Proof in Appendix~\ref{appendix:proof_point_specular}).

\end{highlightbox}
\vspace{-2mm}

Discontinuity is the primary barrier to applying physics-accurate RF models such
as the uniform theory of diffraction (UTD) \cite{kouyoumjian2005uniform,
balanis2012advanced} in a differentiable pipeline. While implicit and
point-based representations such as SDF \cite{chen2024rfcanvas,chen2025radio,park2019deepsdf}
or 3D Gaussian Splatting \cite{kerbl3Dgaussians} offer continuous scene
parameterizations that avoid discontinuity when combined with volume rendering, they smooth out sharp edges that are
essential for preserving physically faithful geometry. So \textit{the discontinuity
challenge arises from modeling such boundaries rather than being a mere artifact
of mesh representation}.

Solutions to the above challenges must produce not only well-defined but also
\textit{physically consistent} gradients. This is crucial for \name, as
physically consistent gradients ensure convergence toward the true scene
parameters whenever the forward model (\ie, ray tracing simulation) is accurate.

\name addresses the challenges in \ttf{two steps}. First,
Sec.~\ref{sec:reparam} reformulates the classical method-of-images ray
tracing \cite{balanis2012advanced} by replacing sequences of discrete triangle
hit tests with a continuous plane-intersection formulation, to ensure that the
mapping from geometry to specular reflection points remains differentiable. Second,
Sec.~\ref{sec:unbiased_diff} replaces the discrete in-triangle validity test
with a physically based weighting term,
ensuring that visibility contributions vary continuously and conserve energy as
specular points approach edges.

\subsection{Reparameterized RF Ray Tracing}
\label{sec:reparam}

In \name, the set of candidate propagation paths is derived analytically from
geometry. 
We model the received electric field $E(\mathbf{p}_\text{rx})$ as the coherent sum over all propagation paths between transmitter $\mathbf{p}_\text{tx}$ and receiver $\mathbf{p}_\text{rx}$.  
Unlike conventional CG methods, our ray tracing method operates under the assumption of dominant specular reflections, yielding a finite and discrete set of deterministic propagation paths:

\vspace{-4mm}
\begin{equation}
\label{eq:e_path}
\small
    E(\mathbf{p}_\text{rx}) = \sum_{\mathcal{P} \in \Omega (\mathbf{p}_\text{tx}, \mathbf{p}_\text{rx})} \mathcal{W}(\mathcal{P}) \cdot A_\mathcal{P} \, e^{-j k L_\mathcal{P}},
\end{equation}
\vspace{-2mm}

\noindent where $\mathcal{P}$ represents an individual propagation \ttf{path} defined as an ordered sequence of interaction points (where the path intersects with triangles in the mesh), and $\Omega(\mathbf{p}_\text{tx}, \mathbf{p}_\text{rx})$ denotes the complete set of all valid paths connecting the Tx-Rx positions. The weighting function $\mathcal{W}(\mathcal{P})$ determines the validity of each path, while $A_\mathcal{P}$ and $L_\mathcal{P}$ represent the amplitude attenuation and total path length, respectively, and $k = 2\pi/\lambda$ is the wavenumber.

To circumvent the measure-zero issue of Monte Carlo sampling, we adopt a \textit{two-stage strategy} (Fig. \ref{fig:ray_tracing}, left). We first gather candidate triangle sequences roughly connecting Tx and Rx by randomly launching rays and selecting those that roughly hit the Rx's nearby region. We then apply our reparameterized version of the method of images to deterministically solve the specular reflection constraints, yielding exact specular paths.

In the conventional method of images, a reflection path abruptly disappears once a ray-triangle hit test fails. In contrast, our reparameterized method deterministically constructs 
\textit{all specular paths via supporting planes},  regardless of their visibility, and preserves them throughout: 

\vspace{-5mm}
\begin{equation}
\small
\begin{aligned}
\mathcal{P}(\mathbf{p}_\text{tx}, \mathbf{p}_\text{rx} \mid \triangle_1,\dots,\triangle_m)
&:= \{\mathbf{p}_\text{tx}, \mathbf{p}_\text{ref}^{(1)}, \dots, \mathbf{p}_\text{ref}^{(m)}, \mathbf{p}_\text{rx}\} \\
= &\,\, \mathcal{T}_{\mathrm{path}}\!\left(\mathbf{p}_\mathrm{tx},\mathbf{p}_\mathrm{rx},\{S(\triangle(v_{i,0},v_{i,1},v_{i,2}))\}_{i=1}^m\right)
\end{aligned}
\end{equation}

This $\mathcal{T}_{\mathrm{path}}$ is a \textit{deterministic} reparameterized
formulation of the method of image, that maps a set of triangles $\triangle_i$ and their vertices to a complete propagation path $\mathcal{P}$.  $\triangle_1$ to $\triangle_m$ are triangle candidates for each path obtained from the first stage path exploration (Appendix ~\ref{appendix:method_of_images}). $S(\cdot)$ denotes the infinite supporting plane of the triangle.  The resulting path
length $L_\mathcal{P}$, amplitude $A_\mathcal{P}$, and phase are all continuous
and differentiable \wrt $\mathbf{p}_\text{tx}$, $\mathbf{p}_\text{rx}$, and
every vertex $v_{i,j}$, as the mapping is defined via intersections with the
infinite planes supporting the triangles.
With our reparameterized formulation, \textit{discontinuities are shifted from the path
construction stage into the weighting function} $\mathcal{W}(\cdot)$. As geometry
changes, energy is redistributed among potential specular paths, rather than
paths being destroyed or created. 

Given a specular path $\mathcal{P}$, conventional methods declare it valid 
when all reflection points lie strictly within the boundary of their
respective triangle with vertices $(v_0, v_1, v_2)$ using barycentric coordinates $(u,w)$ (Appendix~\ref{appendix:method_of_images}):
\vspace{-1.5mm}
\begin{equation}
\small
\begin{aligned}
\mathbf{p} \in \triangle(v_0, v_1, v_2)&:= \left( u \geq 0 \right) \land \left( w \geq 0 \right) \land \left( u + w \leq 1 \right) \\
\mathcal{W}(\mathcal{P}) &:= \bigwedge\limits_{i=1}^m \left( \mathbf{p}_\text{ref}^{(i)} \in \triangle_i \right) {\text{\tiny  \colorbox{gray!10}{ \parbox{2.2cm}{\centering Conventional Method\\  (Discontinuous)}}}}
\end{aligned}
\label{eq:weight_conventioanl}
\end{equation}
\vspace{-4mm}

However, this binary weighting function introduces discontinuities. A straightforward remedy involves replacing the hard boundary constraints with smooth approximations. Specifically, consider a \textit{baseline method} that uses soften triangle function via sigmoid activations:

\begin{figure}[t!]
\centering
\includegraphics[width=0.48\textwidth]{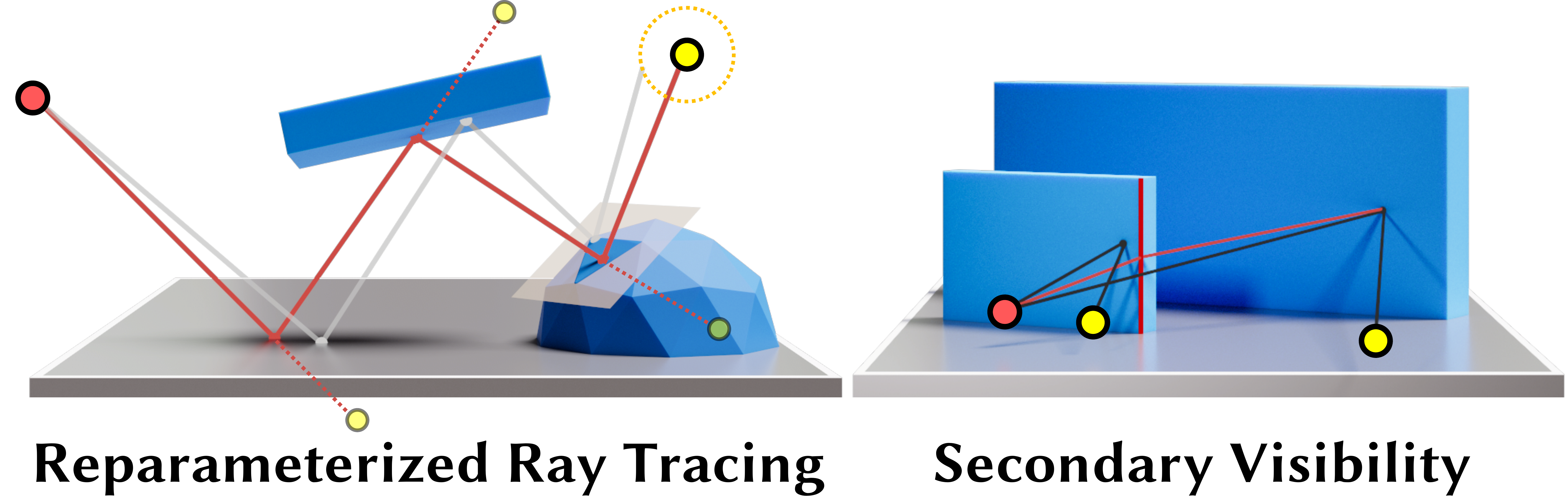}
\put(-243,68){{\small $p_{\text{tx}}$}}
\put(-128,68){{\small $p_{\text{rx}}$}}
\put(-93,20){{\small $p_{\text{tx}}$}}
\put(-76,16){{\small $p_{\text{rx}}^t$}}
\put(-32,16){{\small $p_{\text{rx}}^0$}}
\put(-70,50){{\small $\mathcal{E}$}}
\put(-140,33){{\small $\mathcal{S}(\triangle)$}}
\put(-218,23){{\footnotesize $p_{\text{ref}}^0$}}
\put(-184,45){{\footnotesize $p_{\text{ref}}^1$}}
\put(-168,34){{\footnotesize $p_{\text{ref}}^2$}}
\put(-33,23){\vector(-1,0.067){38}}
\put(-96,40){{\small $\mathcal{S}(\triangle_a)$}}
\put(-50,49){{\small $\mathcal{S}(\triangle_b)$}}
\caption{ Reparametrized RF Ray Tracing and \name  Secondary Visibility. }
\label{fig:ray_tracing}
\end{figure}
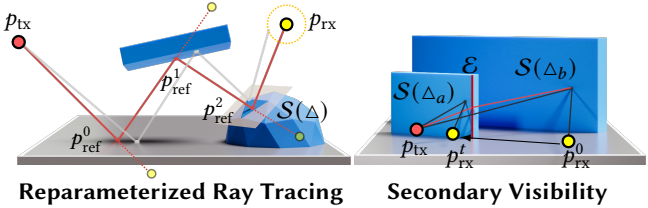

\vspace{-4mm}
\begin{equation}
\small
\begin{aligned}
\mathbf{p} \in_\text{soft} \triangle(v_0, v_1, v_2)&:= \min\left( \sigma(k u),\ \sigma(k w),\ \sigma(k(1 - u - w)) \right) \\
\mathcal{W}_\text{soft}(\mathcal{P}) &:= \prod_{i=1}^m \left( \mathbf{p}_\text{ref}^{(i)} \in_\text{soft} \triangle_i \right) {\text{\tiny  \colorbox{gray!10}{ \parbox{2.2cm}{\centering Baseline Method\\Soften Triangle (Biased)}}}}
\label{eq:soft_triangle}
\end{aligned}
\end{equation}
\vspace{-4mm}

\noindent where $\sigma(\cdot)$ denotes the sigmoid function and $k$ controls the sharpness of the transition. While this formulation enables gradient computation, it allows physically incorrect contributions from paths whose reflection points fall outside the true triangle boundaries. Moreover, the smoothing factor $k$ presents a difficult trade-off: a small $k$ introduces significant physical bias by over-smoothing the geometry, while a large $k$ risks producing unstable gradients. To enable stable RF optimization, a \textit{physically consistent differentiation} is needed.

\subsection{Physically Consistent Differentiation via Edge Diffraction Tracing}
\label{sec:unbiased_diff}
The core challenge towards smooth and correct gradients lies in the path validity function $\mathcal{W}(\mathcal{P})$, which is discontinuous due to two binary checks: \textit{primary visibility} (if reflection points are within triangle boundaries) and \textit{secondary visibility} (if the path is occluded by other objects). We address both issues by smoothing the specular effects through equivalent edge diffraction tracing.

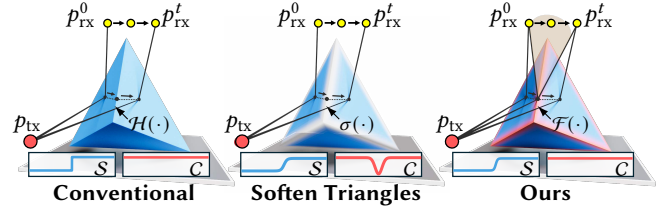
\begin{figure}[t!]
\centering
\includegraphics[width=0.48\textwidth]{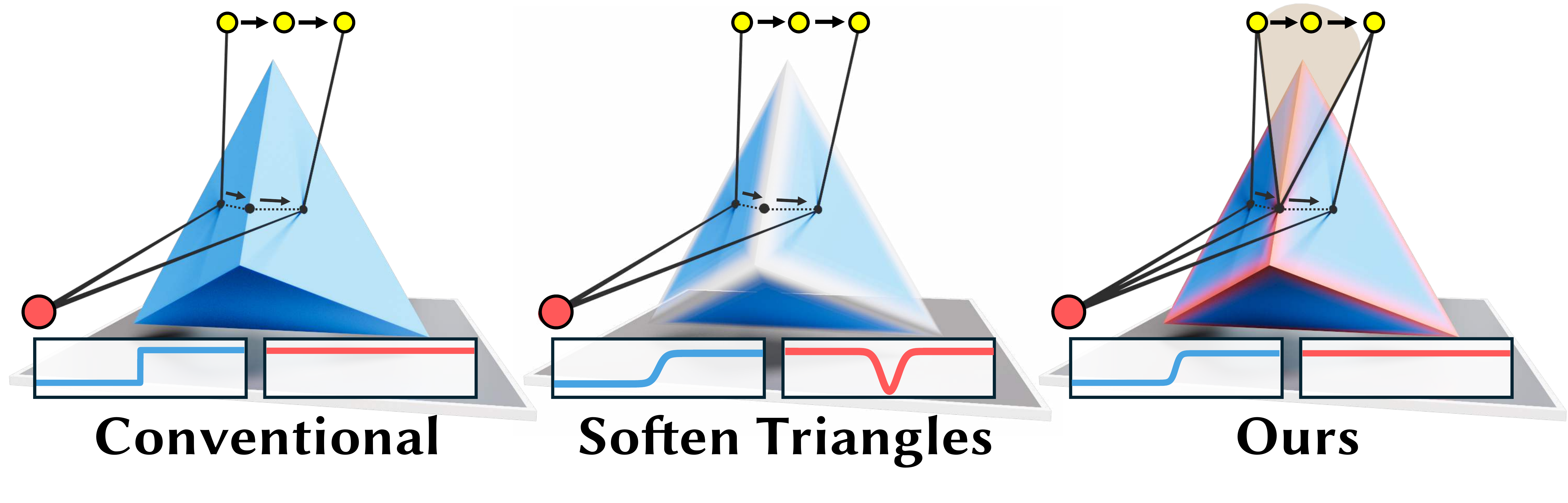}
\put(-243,35){{\small $p_{\text{tx}}$}}
\put(-163,35){{\small $p_{\text{tx}}$}}
\put(-83,35){{\small $p_{\text{tx}}$}}
\put(-223,75){{\small $p_{\text{rx}}^0$}}
\put(-143,75){{\small $p_{\text{rx}}^0$}}
\put(-63,75){{\small $p_{\text{rx}}^0$}}
\put(-187,75){{\small $p_{\text{rx}}^t$}}
\put(-107,75){{\small $p_{\text{rx}}^t$}}
\put(-27,75){{\small $p_{\text{rx}}^t$}}
\put(-200,38){\vector(-1,1){5}}
\put(-200,34){\scriptsize $\mathcal{H}(\cdot)$}
\put(-120,38){\vector(-1,1){5}}
\put(-120,34){\scriptsize $\sigma(\cdot)$}
\put(-40,38){\vector(-1,1){5}}
\put(-40,34){\scriptsize $\mathcal{F}(\cdot)$}
\put(-16,17){\scriptsize $\mathcal{C}$}
\put(-52,17){\scriptsize $\mathcal{S}$}
\put(-96,17){\scriptsize $\mathcal{C}$}
\put(-132,17){\scriptsize $\mathcal{S}$}
\put(-177,17){\scriptsize $\mathcal{C}$}
\put(-213,17){\scriptsize $\mathcal{S}$}
\vspace{-2mm}
\caption{Wedge discontinuity as Rx location translates from $P_\text{rx}^0$ to $P_\text{rx}^t$. $\mathcal{S}$: continuity of changes in signal strength. $\mathcal{C}$: total energy conservation during translation. Conventional: discontinuous step. Softened triangle: heuristic sigmoid, physically incorrect. Ours: continuous and physically correct---As the reflection point approaches the edge, the purely reflection contribution is smoothly redistributed into a diffraction component, as captured by the transition function $\mathcal{F}(\cdot)$.}
\label{fig:unbiased_diff}
\end{figure}

\noindent\textbf{Handling Primary Visibility through a Physically Based Edge Transition Function.}
Instead of treating each triangle face in isolation, we consider adjacent
triangle faces that form a wedge (Fig.~\ref{fig:unbiased_diff}). This wedge
construct applies universally: it can be a true geometric wedge, an artificial
wedge from meshing a smooth surface, or a boundary edge where the second face is
treated as free space. Instead of the conventional step-function test
(Eq.~\eqref{eq:weight_conventioanl}), we model the field contribution as a
superposition of the fields reflected from each triangle face. We use a
\textit{diffraction transition function} $\mathcal{F}(\cdot)$ to create an
energy conserving weight that smoothly transfers contribution across the wedge's
edge. This correctly captures the continuous evolution of the field as the
reflection point traverses the edge, including its smooth decay into the shadow
region at boundaries.

Our goal is \textit{not to simulate diffraction} for its (often small) energy contribution, but to achieve differentiability by exploiting the physical fact that \textit{specular energy continuously transitions into a diffracted component at a wedge}. 
In particular, the valid specular sample on the edge that preserves the law of reflection coincides with the diffraction point, of which there is exactly one per wedge, allowing us to track the existing diffraction path with negligible overhead. This continuity is the geometric property formalized below:

\vspace{-2mm}
\begin{highlightbox}[ Differentiating Specular Paths via Diffraction]
Let $\mathbb{W}$ be a wedge defined by faces $\triangle_a$ and $\triangle_b$ joined at edge $\mathcal{E}$. 
Then there exists a unique point $p^*\in \mathcal{E}$ such that the incident and reflected directions satisfy the law of reflection with respect to the corresponding wedge faces in the limiting sense (Proof in Appendix \ref{appendix:proof_diffraction}):

\vspace{-4mm}
\begin{equation}
\lim_{\epsilon \to 0} \angle(\mathbf{p}_\text{tx} - \mathbf{p}^*_\epsilon, \mathbf{p}_1) = \lim_{\epsilon \to 0} \angle(\mathbf{p}_\text{rx} - \mathbf{p}^*_\epsilon, \mathbf{p}_2),
\end{equation}
\vspace{-4mm}

where $\mathbf{p}^*_\epsilon$ denotes a point approaching the wedge edge from within the adjacent face, and $\mathbf{n}_1, \mathbf{n}_2$ are the face normals. This point $\mathbf{p}^*$ coincides with the edge diffraction point in the Uniform Theory of Diffraction \cite{kouyoumjian2005uniform}.
\end{highlightbox}
\vspace{-2mm}

Mathematically, the specular reflection point on a wedge must satisfy both
\ttf{(i)} the law of reflection (incidence angle equals reflection angle)
and \ttf{(ii)} the geometric constraint that the reflected path lies on the
wedge surfaces. 
This condition yields a unique solution lying exactly on the wedge's edge. In
the classical UTD \cite{kouyoumjian2005uniform,
balanis2012advanced}, this unique edge point is referred to as the
\emph{diffraction point}, which defines the origin of the diffracted field,
and can be modeled as:
\begin{equation}
\small
    \mathbf{E}_\mathrm{diffract}(L) = \mathbf{E}_i(\mathbf{p}_d) \cdot D_{\mathrm{diffract}} \cdot A(L) e^{-jkL}
    \label{eq:diffraction_field}
\end{equation}
\noindent where $\mathbf{E}_i(\mathbf{p}_\mathrm{diffract})$ is the electric field incident on the diffraction point $\mathbf{p}_\mathrm{diffract}$.  
$A(L) = 1/\sqrt{L}$ is the amplitude attenuation factor.
$e^{-jkL}$ is the phase propagation term, where $L=\|\mathbf{r}-\mathbf{p}_d\|$ denotes the propagation path length from the diffraction point $\mathbf{p}_\mathrm{diffract}$ to the observation point $\mathbf{r}$.  
$D_\text{Diffract}$ is the diffraction coefficient of a wedge and can be modeled based on the UTD.

\begin{equation}
\scriptsize
D_{\mathrm{diffract}} = -\frac{e^{-j\pi/4}}{2n\sqrt{2\pi k}} \bigg[ 
 \cot\!\left(\tfrac{\pi - (\phi \mp \phi')}{2n}\right) 
   \colorbox{gray!10}{$\mathcal{F}(kLa^\mp)$}
 + \cot\!\left(\tfrac{\pi + (\phi \mp \phi')}{2n}\right) 
   \colorbox{gray!10}{$\mathcal{F}(kLb^\mp)$}
\bigg],
\end{equation}
\noindent where the $\mp$ notation compactly represents the four cotangent-transition function terms, 
covering $(La^-, La^+)$ and $(Lb^-, Lb^+)$ in the full UTD expression. $n = \pi / \alpha$ defines the wedge sharpness given the interior angle $\alpha$, $k = 2\pi / \lambda$ is the wavenumber, and $\phi, \phi'$ are the observation and incidence angles measured from a fixed edge reference. The distance terms $La^-$ and $La^+$ represent the effective paths from the diffracted point to the receiver via the wedge edges, modulated by $\mathcal{F}(\cdot)$. 

We adopt the transition function $\mathcal{F}(\cdot)$ in UTD and reinterpret it as a physically grounded path validity weight. This function is the key to ensuring physical continuity across shadow boundaries:
\vspace{-2mm}
\begin{equation}
\small
   \mathcal{F}(x) = 2j\sqrt{x} e^{jx} \int_{\sqrt{x}}^{\infty} e^{-j\tau^2} d\tau, 
\label{eq:transition}
\end{equation}

\vspace{-2mm} 
\noindent where $x = kL_a$ is the path length scaled by wavenumber $k$, 
and $L_a$ is the distance from the wedge tip to the observer along the ray. 
The integral $\int_{\sqrt{x}}^\infty e^{-j\tau^2} d\tau$ corresponds to a complex Fresnel integral, 
capturing edge-induced phase curvature and evaluated via standard Fresnel functions.
$\mathcal{F}(x)$ exhibits well-defined asymptotic behaviors (Proof in Appendix \ref{appendix:asymptotic}): For $x \to \infty$ (far from boundaries), $\mathcal{F}(x) \to 1$ recovers the reflection contribution; for $x \to 0$ (near boundaries), $\mathcal{F}(x) \to 0$ suppresses the singular geometric term, enabling proper diffraction-dominated behavior.
This formulation
maintains field continuity and energy conservation as the reflection point approaches/crosses an edge, while eliminating the discontinuities inherent in geometric optics without artificial heuristic smoothing (\eg, sigmoid). 

\noindent\textbf{Handling Multipath Secondary Visibility.}
In multi-path scenarios, a specular segment 
$
\mathcal{P}_a = (\mathbf{p}_{\mathrm{tx}}, \mathbf{p}^{(i)}_{\mathrm{ref}}, \mathbf{p}_{\mathrm{rx}})
$
on surface $\triangle_a$ may be partially or fully shadowed by an unrelated obstacle $\triangle_b$ along its outgoing path (Fig.~\ref{fig:ray_tracing}, Right). We refer to this occlusion as \emph{secondary visibility}, 
In classical geometric optics ray tracing (Sec.~\ref{sec:background}), this is modeled by a binary mask:
$
E^{\mathrm{GO}} = \mathcal{H}(\gamma)\,E_a^{\mathrm{spec}},
$
where $\gamma$ is the signed clearance between the propagation segment $(\mathbf{p}^{(i)}_{\mathrm{ref}},\mathbf{p}_{\mathrm{rx}})$ and the silhouette of $\triangle_b$, and
$E_a^{\mathrm{spec}}$ is the specular field contribution from $\triangle_a$.
This binary masking causes discontinuities at $\gamma=0$.

In our method, the first stage path exploration ensures that all  
specular paths $\mathcal{P}_a$ are represented. For each segment of such a path, 
we perform a BVH-based shadow test against the full scene to identify 
potential blockers and their silhouette edges $\mathcal{E}_c$. 
These edges then provide the corresponding diffraction paths. 
Secondary visibility is handled by smoothly attenuating $E_a^{\mathrm{spec}}$:
$w(\gamma) = \mathcal{F}(k L_a)$,
where $L_a$ is the effective distance from $\mathcal{E}_c$ to the receiver projected along the shadow boundary direction.
The diffraction path
$
\mathcal{P}_d = (\mathbf{p}_{\mathrm{tx}}, \mathbf{p}_d, \mathbf{p}_{\mathrm{rx}}), $ where $\mathbf{p}_d \in \mathcal{E}_c
$
contributes
a diffraction field given by Eq.~\eqref{eq:diffraction_field}, 
with $s = L_d$ denoting the total path length via $\mathbf{p}_d$.
The total field for this segment is then:

\begin{equation}
    E_{\mathrm{total}} = w(\gamma)\,E_a^{\mathrm{spec}} + E_{\mathrm{diffract}},
    \label{eq:e_total}
\end{equation}

\noindent again ensuring that as $\gamma \to 0$, the specular term decays
continuously while the diffraction term rises, thus preserving field continuity. This $E_{\mathrm{total}}$ represents the complete contribution of this specific path $P$ to the summation in Eq.~\eqref{eq:e_path}.

\noindent\textbf{Differentiation.}
To compute the gradient of the total received field $E(p_\text{rx})$ \wrt a mesh vertex $v$, we apply the chain rule. A vertex $v$ defines the geometry of the triangles it belongs to. A change in $v$ perturbs the reflection and diffraction paths that interact with these triangles, thereby altering their lengths, amplitudes, and phase contributions. The gradient is expressed as a sum over all affected paths $\mathcal{P} \in \Omega$:

\begin{equation}
\small
    \frac{\partial E(p_\text{rx})}{\partial v} = \sum_{\mathcal{P} \in \Omega} \frac{\partial}{\partial v} \left( \mathcal{W}(\mathcal{P}) \cdot A_\mathcal{P} \, e^{-j k L_\mathcal{P}} \right).
\end{equation}

Our key contribution here is a physically accurate, differentiable formulation of the edge-weighting term $\mathcal{W}(\mathcal{P})$:

\begin{equation}
\small
\mathcal{W}_\text{\name}(\mathcal{\mathcal{P}}) := \prod_{i=1}^m \left( \mathcal{F}(kL_a(\alpha))   \right),
\end{equation} 
$\mathcal{W}_\text{\name}(\cdot)$ generalizes the principle from Eq.~\eqref{eq:e_total}. It multiplicatively applies the transition function $\mathcal{F}(\cdot)$ across all $m$ path interactions to yield the total weight.

For a path $\mathcal{P}$ interacting with an edge $\mathcal{E}$ formed by triangles $\triangle_a$ and $\triangle_b$, the direct differentiation would incorporate the binary Heaviside mask:

\begin{equation}
    E_{\mathcal{E}}^\text{dire} =  \mathcal{H}(\alpha(v_0, v_1)) E_a + \mathcal{H}(-\alpha(v_0, v_1)) E_b,
\end{equation}

\noindent which causes discontinuities when the reflection point crosses the edge boundary.  
The soften triangle method in Eq ~\eqref{eq:soft_triangle} replaces the hard step with sigmoid smoothing:

\begin{equation}
    E_{\mathcal{E}}^\text{soft} =  \sigma(\triangle_a) E_a + \sigma(\triangle_b) E_b,
\label{eq:sigmoid_weight}
\end{equation}

\noindent ensuring continuity but losing physical precision.  
We instead use the diffraction transition function to obtain a 
physically-consistent and energy-conserving transition:

\begin{equation}
    E_{\mathcal{E}}^\text{\name} = \mathcal{F}(x_a) E_a + \mathcal{F}(x_b) E_b,
\end{equation}

\noindent where $E_a$ and $E_b$ are the complex field contributions from the specular paths reflecting off the infinite planes of $\triangle_a$ and $\triangle_b$, respectively. The arguments $x_a = k L a_a$ and $x_b = k L a_b$ in the UTD formulation involve geometric factors $a_a$ and $a_b$, which represent the normalized angular distances to the reflection and edge boundaries of each face. 

The gradient with respect to vertex $v$ is then obtained by applying the product and chain rules:
\begin{equation}
\footnotesize
\label{eq:gradient_wedge}
    \frac{\partial E_{\mathcal{E}}}{\partial v} = \left( \frac{\partial \mathcal{F}(x_a)}{\partial v} E_a + \mathcal{F}(x_a) \frac{\partial E_a}{\partial v} \right) + \left( \frac{\partial \mathcal{F}(x_b)}{\partial v} E_b + \mathcal{F}(x_b) \frac{\partial E_b}{\partial v} \right)
\end{equation}

Each term in Eq.~\eqref{eq:gradient_wedge} is now differentiable. The gradient of the transition function $\mathcal{F}$ is given by:
\begin{equation}
\small
    \frac{\partial \mathcal{F}(x)}{\partial v} = \mathcal{F}'(x) \frac{\partial x}{\partial v} = \mathcal{F}'(x) \frac{\partial (kLa)}{\partial v},
\end{equation}
where the derivative of the geometric term $kLa$ encapsulates the change in path geometry due to the vertex perturbation $v$.

The remaining terms, $\frac{\partial E_a}{\partial v}$ and $\frac{\partial E_b}{\partial v}$, account for the change in the specular path contributions themselves. For a generic path term $E_p = A_p e^{-j k L_p}$, the gradient is:
\begin{equation}
\small
    \frac{\partial E_\mathcal{P}}{\partial v} = \left( \frac{\partial A_\mathcal{P}}{\partial v} - j k A_p \frac{\partial L_p}{\partial v} \right) e^{-j k L_\mathcal{P}}
\end{equation}
The gradients of path length $L_p$, amplitude $A_p$, and the diffraction parameter $La$ are computed by differentiating the geometric constructs of the reparameterized method of images (Sec.~\ref{sec:reparam}).
Since these operations are compositions of vector additions, dot products, and cross products, they are differentiable with gradients that are well-defined and computable via automatic differentiation. This chain of differentiable geometric calculations provides the final gradients required to update the vertex positions $v$, enabling physically correct, gradient-based optimization of the scene geometry.

\label{sec:surrogate}
\begin{figure}[t!]
\centering
\includegraphics[width=0.48\textwidth]{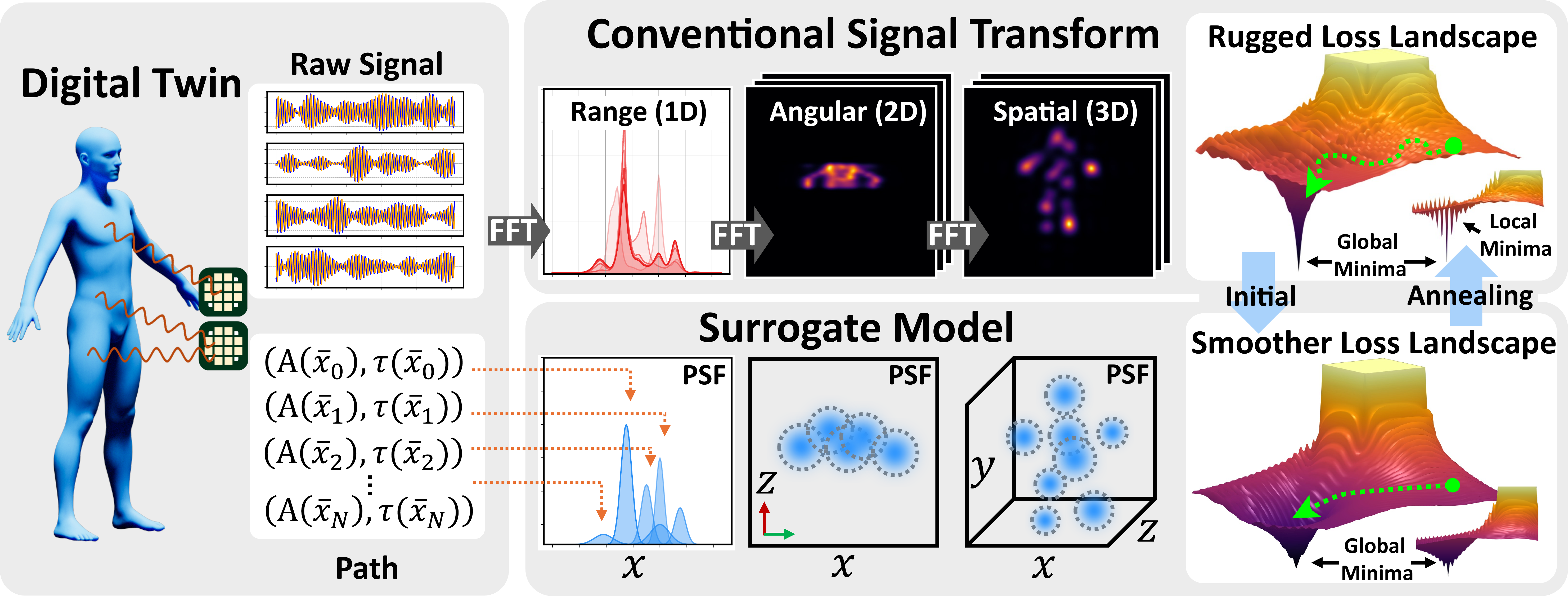}

\caption{\name's surrogate model provides a smoother loss landscape for faster convergence in the early stages, while progressively annealing to the conventional signal-domain transform for unbiased results. }
\label{fig:surrogate}
\end{figure}

\section{Surrogate Model for Signal-Domain Transform}

\textbf{Root causes of non-convexity.} 
RF systems estimate range and angle from modulated signals, but both are subject to severe non-convexities arising from the periodic nature of electromagnetic waves and multipath interference. Due to the inherent $2\pi$ phase periodicity, multiple path lengths can yield the same phase relative to the source. Moreover, paths with a close-to-$\pi$ phase offset interfere destructively, creating sharp valleys in received power and unstable gradients.  Similarly, different multipath components can align in angular space, producing sidelobes and ambiguities that lead to spurious local minima in optimization. 

Wideband radar waveforms such as frequency-modulated continuous wave (FMCW) or orthogonal frequency division modulation (OFDM) \cite{sit2011ofdm} reduce phase ambiguity by spreading energy across multiple frequency channels, allowing delays of different paths to be distinguished in the Fourier domain. However, with finite bandwidth and a limited number of samples, Fourier analysis still suffers from limited resolution and sidelobes~\cite{Oppenheim1998digital}, leaving residual ambiguities and local minima when transforming ray-traced multipath responses into the signal domain. In the context of differentiable RF simulation, these artifacts manifest as misleading gradients during geometry or material optimization.  
High-resolution spectral estimation techniques such as MUSIC \cite{schmidt1986multiple} can suppress sidelobes, but they rely on eigenvalue decomposition of the sample covariance matrix. While differentiable in principle, such operations suffer from extremely high gradient variance and are therefore impractical for stable gradient-based optimization (Appendix \ref{appendix:music}). 

\begin{figure}[t]
\centering

\includegraphics[width=0.49\textwidth]{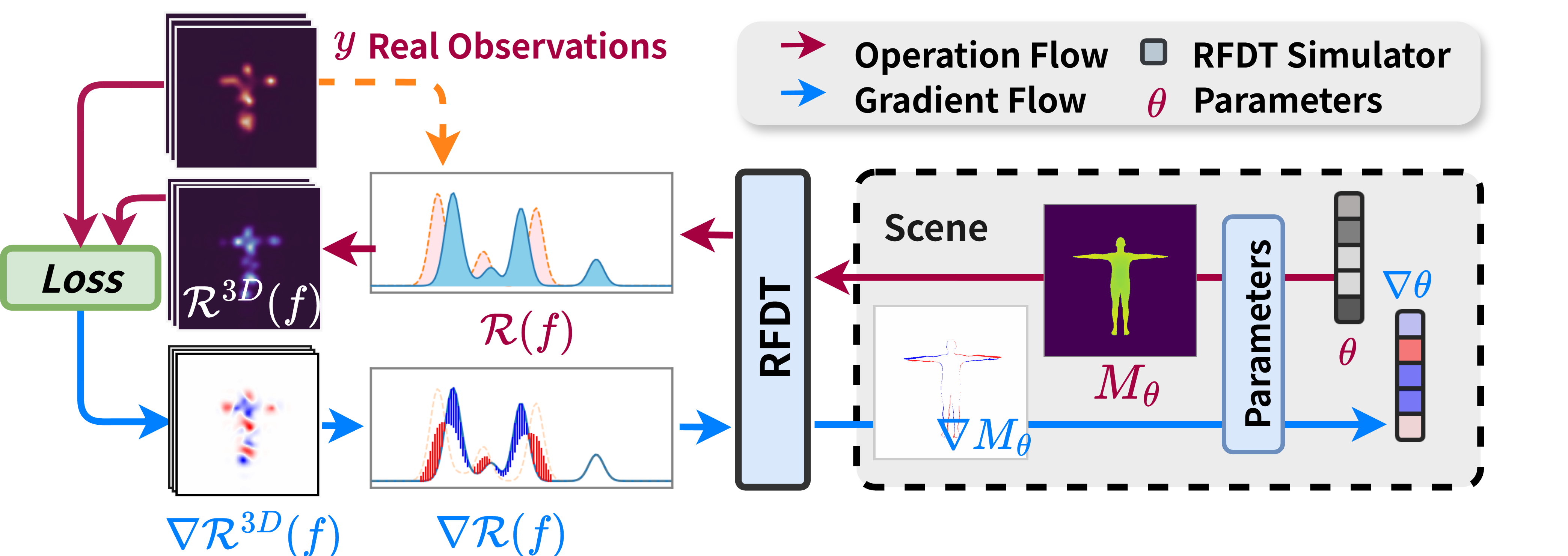}

\caption{ \name forward simulation and backward propagation flow. }
\label{fig:gradient_flow}
\end{figure}

\textbf{Coarse-to-fine surrogate model.}
To address these local non-convexities, we propose a \textit{coarse-to-fine surrogate model} that produces a biased but smooth approximation early in optimization, and progressively anneals to the exact FFT-based transform (Fig.~\ref{fig:surrogate}). This strategy avoids early convergence to spurious minima while still ensuring physical fidelity at convergence.
Specifically, we use a rectangular-window \textit{Point Spread Function}
(PSF) $\mathcal{R}(f)$ \cite{ho2024differentiable,li2024high}, based on the \textit{Dirichlet kernel}, to approximate
the FFT response from each CIR path:

\begin{equation}
    \small
\mathcal{R}(f) = \sum_{i} G_i(\tau_i, f) \cdot \alpha_i, 
\vspace{-2mm}
\end{equation}
This design choice is physically motivated.
For a finite-length sinusoid observed through a rectangular window, the squared FFT magnitude is exactly described by a Dirichlet kernel \cite{lejeune1829convergence,stein2011fourier}. Since each multipath component corresponds to such a sinusoid, the kernel faithfully reproduces both the main-lobe and sidelobe structure of the FFT. This makes it a principled surrogate, in contrast to generic smoothing functions (\eg, Gaussians). The kernel is defined as:
\vspace{-2mm}
\begin{equation}
    \small
G_i(\tau_i, f) = \left|\frac{\sin(N \pi r_i / 2)}{N \sin(\pi r_i / 2)}\right|^2, \quad r_i = \frac{\tau_i - D(f)}{\sigma},
\end{equation}
\noindent where $D(f)$ maps frequency bins to corresponding delay values, $N$ is the FFT size, and $\sigma$ controls the frequency resolution. For the limiting case where $r_i \to 0$, we set $G_i = 1$.

\noindent\textbf{Phase-agnostic initial optimization:} 
During the early stages of the \name optimization (Sec.~\ref{sec:DTopt}), we deliberately omit phase information ($\phi_i$) to produce a smoother, more convex loss landscape. This approximation trades short-term fidelity for improved convergence stability.  

\noindent\textbf{Progressive transition to exact FFT processing:}  
As optimization proceeds, we gradually anneal from the smooth surrogate $R(f)^s$ to the exact FFT-based profile $R_{\text{FFT}}(f)$ by linearly blending the two:  
\begin{equation}
    \small
R(f) = \lambda(t) \cdot R_{\text{FFT}}(f) + \big(1-\lambda(t)\big) \cdot R(f)^s,
\end{equation}
where $\lambda(t)\in[0,1]$ follows a monotone annealing schedule from $0$ to $1$.  
\revise{This coarse-to-fine smoothing perturbs early gradients only within a sub-wavelength phase tolerance to mitigate phase-level non-convexities. After the transition to the exact signal-transform model, all schedules converge to the same optimum, indicating negligible sensitivity to the initial smoothing bias.}

The surrogate $R(f)^s$ remains analytically differentiable. The partial derivatives of the Dirichlet kernels are:
\begin{equation}
    \footnotesize
    \frac{\partial G_i}{\partial \tau_i} 
    = \frac{2G_i}{\sigma} \cdot \frac{\pi}{2} \cdot \left[\frac{N \cos(N\pi r_i/2)}{\sin(\pi r_i/2)} - \frac{\cos(\pi r_i/2) \sin(N\pi r_i/2)}{\sin^2(\pi r_i/2)}\right],
\end{equation}
leading to gradients of the surrogate response \wrt each CIR component:
\vspace{-4mm}
\begin{equation}
\small
    \frac{\partial R(f)^s}{\partial \tau_i} = \frac{\partial G_i}{\partial \tau_i} \cdot \alpha_i, \quad
    \frac{\partial R(f)^s}{\partial \alpha_i} = G_i.
\end{equation}
\vspace{-4mm}

\vspace{-5pt}
\section{\name System Integration}
\label{sec:integration}

With the physically consistent differentiation and the signal-domain surrogate
model in place, gradients can be propagated through every stage of RF simulation
back to the DT parameters. This enables an end-to-end optimization loop, as
illustrated in Fig.~\ref{fig:gradient_flow}. We now describe how the DT is
parameterized, optimized, and implemented.  

\vspace{-4pt}
\subsection{Digital Twin Parameterization}
\label{sec:DTparam}

We represent the DT by a flexible parameter set $\Theta$ encompassing any variables required by downstream use cases.  
These may include geometric descriptors $M_\theta$ (\eg, mesh, voxels)
rigid transformations (translation, rotation, scaling), material properties (\eg, reflection coefficients, dielectric constants; see Appendix~\ref{appendix:material_nlos}), sensor parameters (\eg, radar configuration), and other task-specific quantities.

The DT can be parameterized to deeply integrate with traditional neural network-based methods in two ways.

\begin{figure}[t]
\centering
\includegraphics[width=0.49\textwidth]{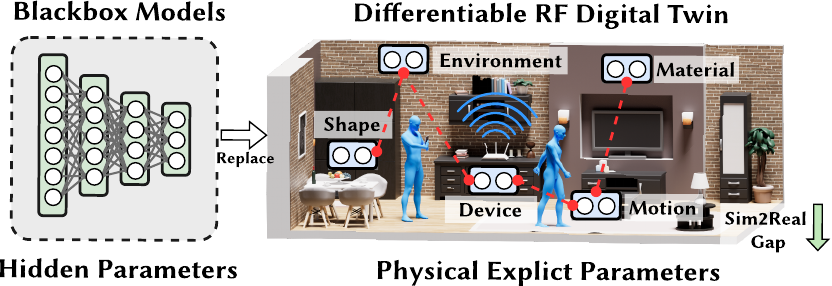}
\vspace{-13pt}
\caption{Replacing Neural Representation with DT. }
\label{fig:whitebox}
\end{figure} 

\noindent\textbf{Neural Network Replacement.}
Unlike conventional neural networks, whose parameters (\eg, weights and biases) are abstract and lack direct physical interpretation, every DT parameter in \name is \emph{explicit} and \emph{physically meaningful}. 
For instance, as shown in Fig.~\ref{fig:whitebox}, a matrix multiplication in a neural network corresponds to a physically based ray-tracing or wave-propagation operator in our DT. Similarly, a learned scalar weight in a multilayer perceptron (MLP) could represent a material property (\eg, dielectric constant) or a geometric attribute (\eg, surface normal) in the DT.
This explicit physical grounding improves interpretability and allows the DT to serve as a trainable physics model for wireless system design. 
When scene and material parameters are made explicit and learnable, few-shot calibration with real measurements can adapt the DT to physical scenes, thereby efficiently narrowing the sim-to-real gap.

\begin{figure}[t]
\centering
\includegraphics[width=\linewidth]{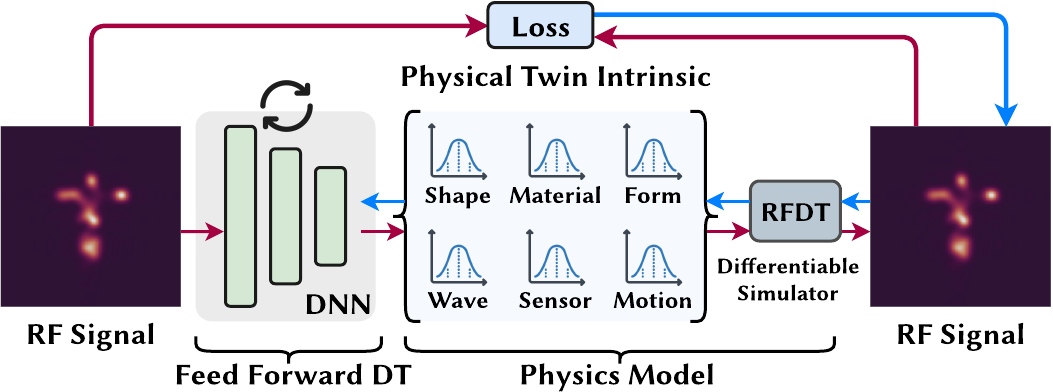}%
\vspace*{-2mm}
\caption{Enhancing Neural Network with DT.}
\label{fig:pinn}
\end{figure}

\noindent\textbf{Neural Network Enhancement.}
The same parameterization also allows seamless integration with neural networks. As illustrated in Fig.~\ref{fig:pinn}, for a Physics-Informed Neural Network (PINN)-like setup,
the DT can act as a differentiable physics module that enforces self-supervised consistency, contributing a physics-based loss term without requiring labels. 
The DT can also be embedded directly into a neural network by treating its physical parameters as learnable weights (\emph{Feed Forward DT}), allowing direct inference of scene parameters from RF observations. This replaces iterative optimization with a single forward pass, enabling runtime parameter estimation while preserving interpretability.  
Our case study (Sec.~\ref{sec:usecase1}) will further verify such advantages.

\vspace{-5pt}
\subsection{Digital Twin Optimization }
\label{sec:DTopt}

We formulate the interaction between the DTs and physical entities as an iterative optimization process, as illustrated in Fig.~\ref{fig:gradient_flow}.
Starting from an initialization $\bm{\theta}_0$, the DT parameters are updated to minimize the discrepancy between simulated and observed RF signals:
\vspace{-2mm}
\begin{equation}
\small
\bm{\theta}^* = \arg\min_{\bm{\theta} \in \Theta} \mathcal{L}(RFDT(\bm{\theta}), \bm{y}),
\end{equation}
\noindent where $\mathcal{L}(\cdot,\cdot)$ is the loss function, and 
\revise{
$\bm{y}$ is the stacked vector of all available measurements, whose dimension depends on the number of observations used in the joint optimization.
}
which can be either real measured RF data or a task-defined target distribution.
We employ a multiscale MSE loss to capture spectral structure at different resolutions, which helps to mitigate the sparsity of the RF spectrum.  

To solve the optimization problem, we use the Adam optimizer with Laplacian regularization:
\begin{equation}
\small
\bm{\theta}_{t+1} \leftarrow \text{Adam}\!\left(\bm{\theta}_{t}, 
   \nabla_{\bm{\theta}} \mathcal{L}(S(\bm{\theta}), \bm{y}) + \beta \mathbf{L}\bm{\theta} \right),
\end{equation}
where $\mathbf{L}\in\mathbb{R}^{n\times n}$ is a sparse, symmetric positive definite Laplacian matrix that discretizes the Dirichlet energy~\cite{solomon2014laplace}, and $\beta$ controls the regularization strength. This prevents overfitting to noisy gradients and encourages smooth updates.
Using regularized Adam, we iteratively refine the DT by sampling subsets of RF
signals, computing the discrepancy gradients and updating the DT parameters.
Upon convergence, the RF simulation output should closely match the observed RF
signals. Meanwhile, \name produces an accurate,  parameterized digital twin
$\bm{\theta}^*$ of the physical scene. 

\vspace{-5pt}
\subsection{Software Implementation}
\label{sec:implementation}
We have completed a full-fledged implementation of \name, including the forward ray tracing pipeline, the differentiation mechanisms (Sec.~\ref{sec:diffRT} and \ref{sec:surrogate}), 
and the DT optimization framework (Sec.~\ref{sec:DTparam} and \ref{sec:DTopt}).
Built entirely from scratch with about 34,000 lines of optimized Python, C++, and CUDA
code, 
our \name engine is not an adaptation of existing frameworks (\eg, Sionna). 
It achieves significantly higher efficiency in both communication and sensing tasks and is explicitly tailored to dynamic scenes as well as monostatic and bistatic radar configurations.

The ray tracing simulation is powered by hardware acceleration through NVIDIA's
OptiX library \cite{parker2010optix} on RT cores \cite{nvidia_ray_tracing}. We
employ a bounding volume hierarchy (BVH)\cite{wald2006building} as the
acceleration structure, reducing complexity to $O(M \log N)$ for $M$ rays and
$N$ triangles. 
We support Doppler and scene dynamics by computing path-specific frequency shifts via velocity projections onto ray directions, and by selectively refitting only the BVH regions associated with moving objects (Appendix~\ref{appendix:doppler}), which avoids the prohibitive cost of full BVH rebuilds under mobility.
The DT optimization seamlessly incorporates the forward
simulator as a differentiable operator. The differentiation is primarily handled
via Dr.JIT \cite{jakob2022dr}, which generates backward functions at runtime.
However, the auto-generated functions may not be optimal
\cite{bangaru2023slang}, resulting in a large computation graph that becomes a
bottleneck in the backpropagation. Therefore, we manually implement custom
backward passes for path differentiation (Sec.~\ref{sec:unbiased_diff}) and the
surrogate model (Sec.~\ref{sec:surrogate}) to optimize its efficiency.

\section{System Evaluation}
\label{sec:eval}

\begin{figure}[t!]
\centering
\rotatebox{-90}{\includegraphics[height=\linewidth]{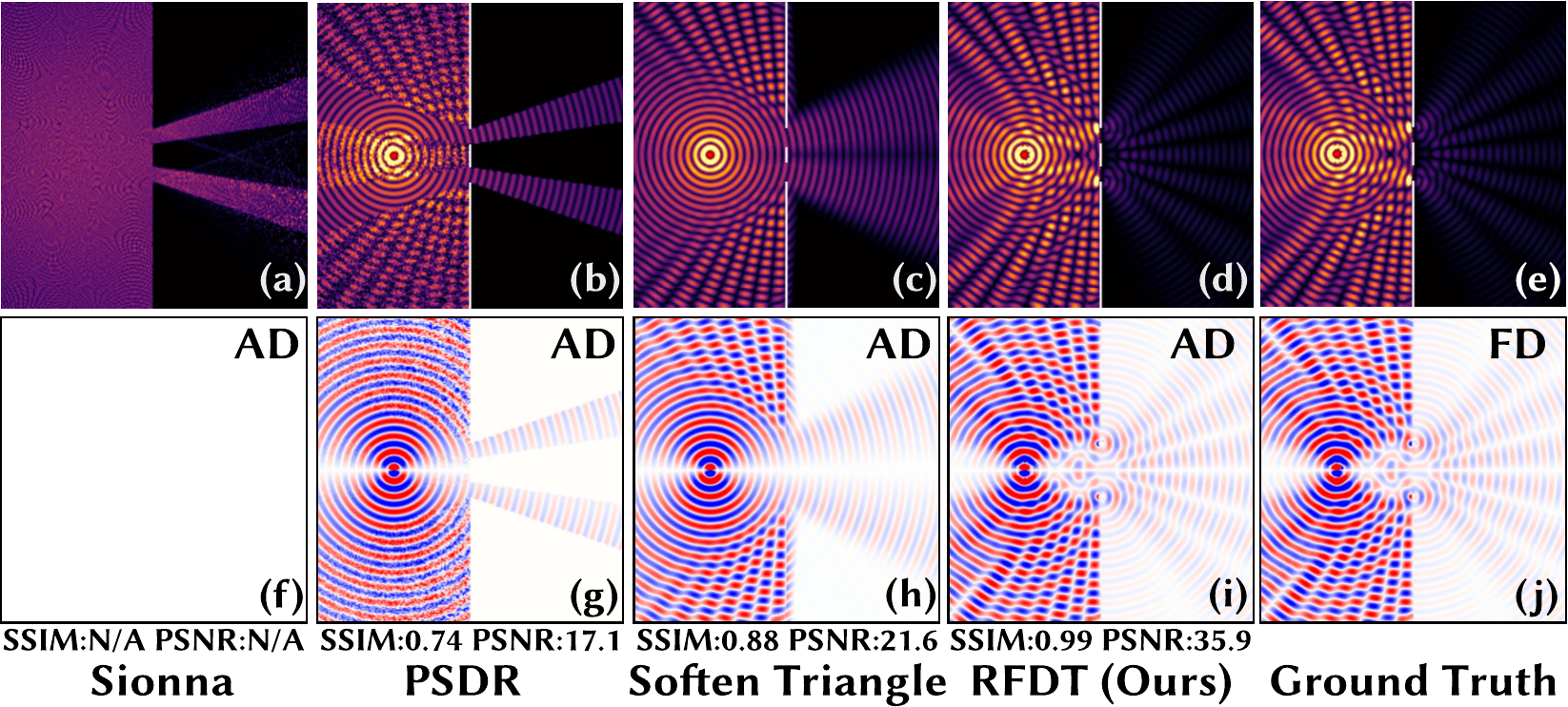}}
\caption{Field simulation (Real part, top) and gradient estimation (bottom) \wrt scene parameter $\theta$.}
\label{fig:grad}
\end{figure} 

\subsection{Verification of Differentiability}
\label{sec:eval:diff}
We first evaluate the effectiveness of \name in gradient estimation. As a
controlled setup (Fig.~\ref{fig:grad}), we use the classic double-slit
interference scene with a carrier frequency of 5~GHz. The relatively long
wavelength makes phase variations easier to interpret in visualizations, though
our method applies across a wide range of carrier frequencies. A target
parameter $\theta$ controls the vertical translation of the Tx source. 
We assign $\theta$ a tangent value of 1 and compute its gradients via PyTorch
forward-mode automatic differentiation \cite{pytorch_forward_ad_beta} for
both the baseline and \name.  

\noindent\textbf{Ground Truth.}
We adopt the finite-difference (FD) method as the ground truth, following common
practice in differentiable rendering and simulation \cite{leveque1998finite,
loubet2019reparameterizing, li2018differentiable}. FD estimates gradients as  
$
\frac{\partial f}{\partial x} \approx \frac{f(x+\epsilon) - f(x)}{\epsilon},
$  
where a small perturbation $\epsilon$ is applied. 
While accurate, FD can only compute one gradient component, making it
impractical for high-dimensional problems where thousands of parameters must be
updated simultaneously. As a result, FD is typically employed only in small,
controlled simulation environments.

\begin{figure}[t!]
\centering
\includegraphics[width=0.49\textwidth]{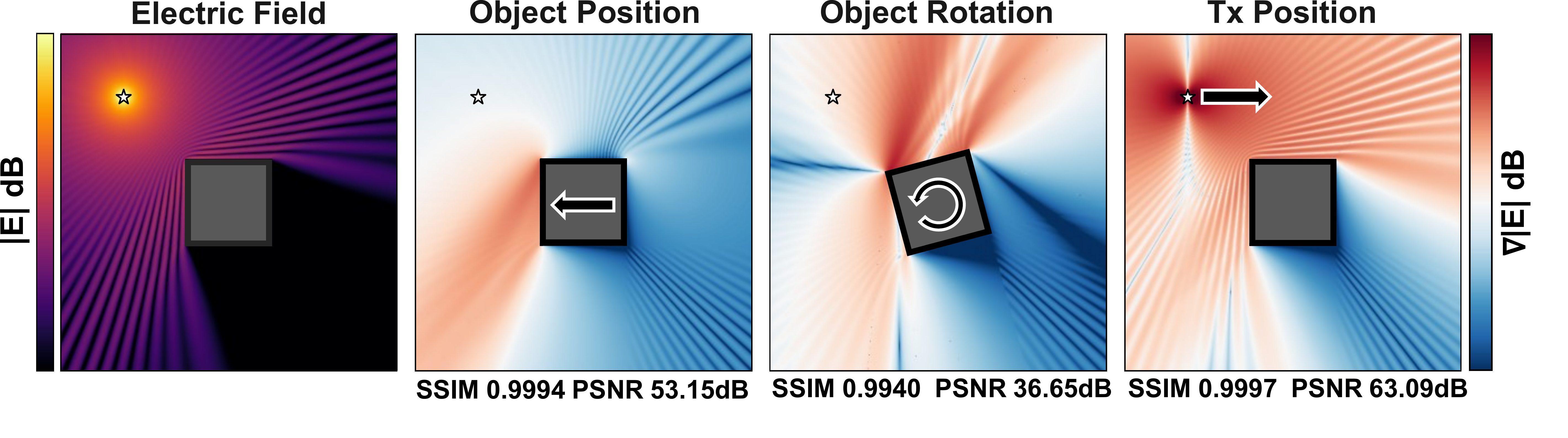}
\caption{\revisetwo{Gradients of the field magnitude \wrt different scene parameters (object position, rotation, and Tx position)}.  
}
\label{fig:grad2}
\end{figure} 

\noindent\textbf{Baseline Comparisons.}
We benchmark against three alternative designs: 
\ttf{(i)} Adapting path-space differentiable rendering (PSDR) \cite{Zhang:2020:PSDR} for differentiable RF simulation. PSDR is a state-of-the-art differentiable rendering method in CG. We use PSDR for the RF 
\cite{chen2025invertwin}. 
\ttf{(ii)} The softened triangle method (Eq.~\eqref{eq:soft_triangle}). 
\ttf{(iii)} NVIDIA's Sionna Simulator. At submission time, Sionna exists in two versions: v1.10 (without diffraction) and v0.19 (with diffraction). We use v0.19 in experiments, and report v1.10 results in the Appendix \ref{appendix:sionna}. 
For all methods, we compute the structural similarity index (SSIM) and peak-signal-to-noise ratio (PSNR) between their
simulated field distributions and the FD ground truth. 
\revise{Note that SSIM prioritizes structural fidelity, whereas PSNR is sensitive to amplitude accuracy. Thus, baselines with imperfect diffraction may yield high SSIM due to correct spatial layouts, despite low PSNR caused by phase and amplitude biases.}

From the results in Fig.~\ref{fig:grad}, we observe that 
PSDR achieves a reasonable SSIM of \textbf{0.8534} and PSNR of \textbf{17.09~dB}, but this is only because it assumes 
all materials have a small nonzero roughness instead of being specular.
It ignores measure-zero specular paths (Sec.~\ref{sec:discontinuity}) and focuses entirely on the scattering effect.  
By doing so, it smooths angular discontinuities but distorts phase accuracy, as
revealed by the noisy and disturbed gradients in Fig.~\ref{fig:grad}(h). 
Sionna's gradients are ill-defined as it is not differentiable \wrt to geometric scene parameters.  
Sionna's forward simulation of radio coverage (Fig.~\ref{fig:grad}(a)) does not account for complex-valued coherent interference but instead applies simplified non-coherent power summation (Appendix~\ref{appendix:sionna}), leading to inaccurate simulation results. 
The softened triangle method achieves an SSIM of \textbf{0.8785} and PSNR of \textbf{21.60~dB}, yet its geometric-level approximation cannot reproduce sharp edges and suffers from energy leakage. In contrast, \name produces results that are nearly identical to the ground truth (SSIM: \textbf{0.9908}; PSNR: \textbf{35.94~dB}).

\revisetwo{Fig.~\ref{fig:grad2} illustrates the evaluation of \name's gradients \wrt different geometric parameters. \name yields results nearly identical to the ground truth across key parameters, including object position, rotation, and transmitter location (Position: SSIM 0.9994, PSNR 53.15dB; Rotation: SSIM 0.9940, PSNR 36.65dB; Tx: SSIM 0.9997, PSNR 63.09dB).  }

\begin{figure}[t!]
\centering
\includegraphics[width=0.48\textwidth]{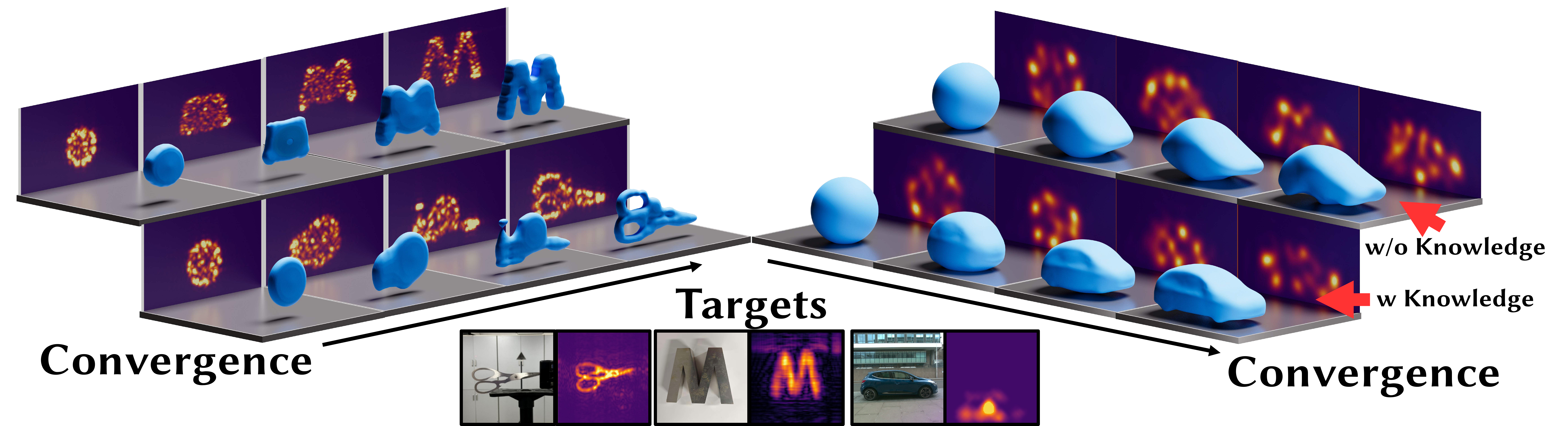}
\caption{Snapshots of DT optimization process. Left: reconstruction of metal desktop objects using the Cascade radar; Right: reconstruction of a real-world vehicle using the Vayyar radar, without and with pre-trained knowledge of ``possible shapes of vehicles''.}
\label{fig:opt}
\end{figure}

\subsection{Convergence and Accuracy of DT Reconstruction} 
\label{sec:convergeEval}
Next, we evaluate \name on a 3D reconstruction task using real signals collected
by a 60GHz Vayyar TrigB imaging radar \cite{Vayyar_2024} with a 20T$\times$20R
MIMO antenna \cite{AWR1843BOOST}. The radar was co-located with an Intel
Realsense D455 RGBD camera \cite{intel_realsense} to provide ground-truth 3D
shapes  (Appendix ~\ref{appendix:setup}). Additionally, we employ the radar signal dataset from \cite{li2024high},
collected by a 77GHz TI MMWCAS (Cascade) radar (12T$\times$16R MIMO
antenna) with various 3D objects placed in front. By default, we initialize the
scene geometry as a sphere with metal materials (according to ITU database
\cite{ITU-R-P2040}), and then jointly optimize geometry and material properties
following Sec.~\ref{sec:DTopt}.  

\begin{figure}[t!]
\centering
\includegraphics[width=0.48\textwidth]{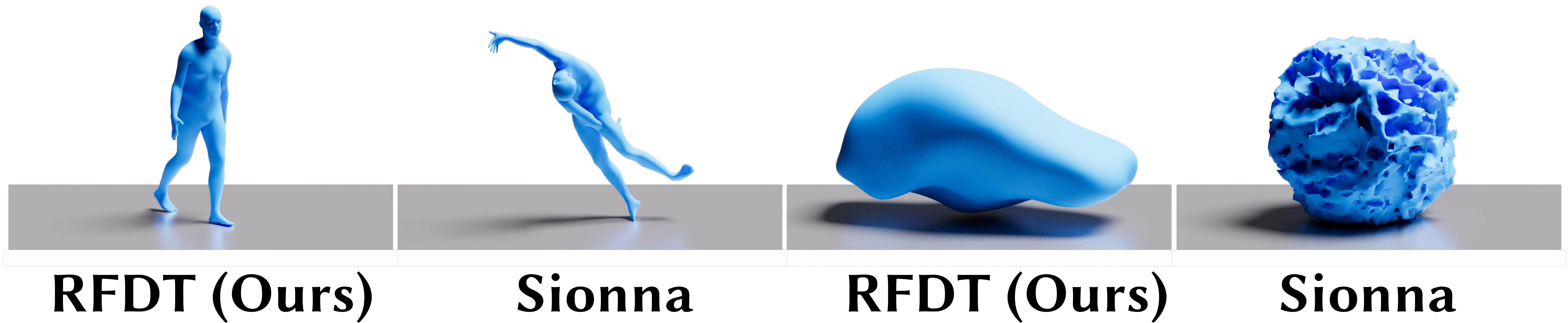}
\caption{Reconstructing human posture and unknown object shape using radar signals. 
}
\label{fig:ill}
\end{figure}

\begin{figure}[t!]
\centering
\includegraphics[width=0.45\textwidth]{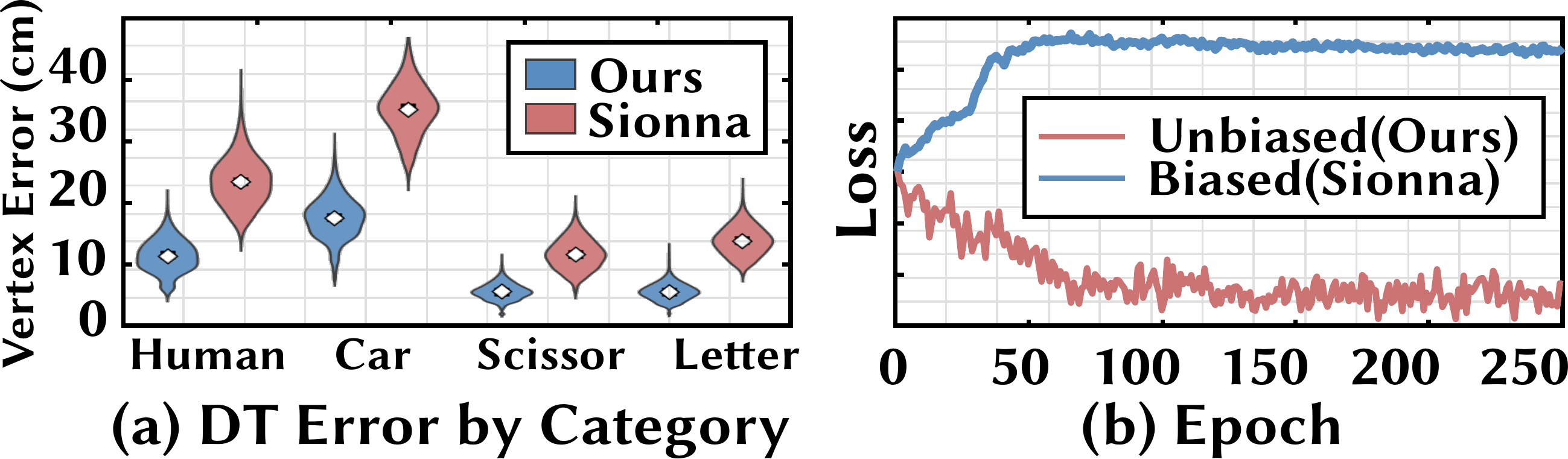}
\caption{ (a) Mean vertices error of DT reconstruction; (a) Loss history of the convergence for human posture.}
\label{fig:system_result_a}
\end{figure}

Fig.~\ref{fig:opt} shows snapshots of the reconstruction process for desktop objects and a car. 
Fig.~\ref{fig:ill} further showcases the reconstruction results for human and car. 
\name achieves a median error of 15.4~cm and an MAE of 16.7~cm when reconstructing a car, and a
median error of 1.7~cm and MAE of 2.4~cm when reconstructing a metal scissor and
metallic letter. 
The distribution of vertex errors across object categories is shown in Fig.~\ref{fig:system_result_a}(a). These results indicate that \name correctly reconstructs the target's shape and pose with low vertex error by properly handling geometric differentiation, whereas the corrupted meshes produced by Sionna result in high vertex error and variation. 
Fig.\ref{fig:system_result_a}(b) shows the loss over epochs. \name's DT
optimization converges within 300 rounds, taking only 3.5 minutes in total on an 
NVIDIA A6000 GPU. 
In contrast, Sionna failed to converge (also indicated in the two examples in Fig.~\ref{fig:ill}). 
This failure roots in its inability to capture gradients at geometric boundaries, leading to corrupted gradients, causing divergence during optimization and random/corrupted reconstruction. 

\begin{figure}[t!]

\centering
\includegraphics[width=0.48\textwidth]{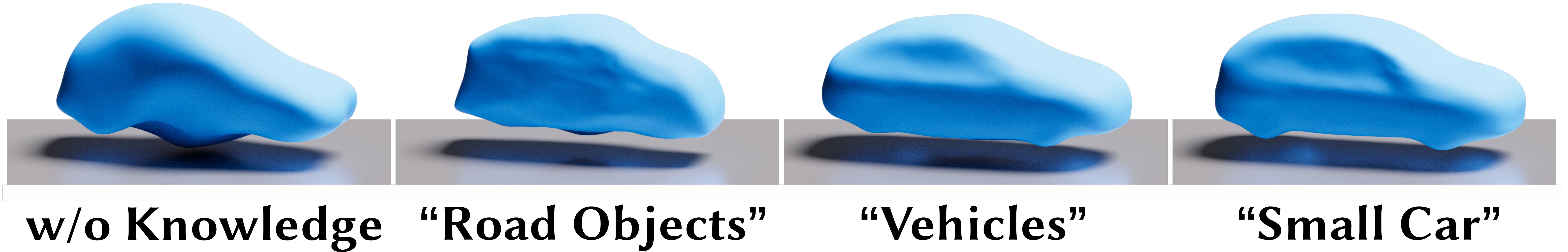}
\caption{Impact of the granularity of knowledge provided to \name when performing shape reconstruction.}
\label{fig:prior_knowledge}
\end{figure}

\noindent\textbf{Impact of prior knowledge in DT optimization.}
We follow Sec.~\ref{sec:DTopt} to embed three different levels of knowledge to \name to facilitate the car reconstruction task: ``Small Car'', embedding 3D models of sedans, SUVs, minivans, hatchbacks, and station wagons from ShapNet; ``Vehicles'', embedding cars, buses, trucks, trains, and motorcycles; and ``Road Objects'', embedding all types of vehicles, road signs, benches, and trash bins. 
The right side of Fig.~\ref{fig:prior_knowledge} shows that even moderate knowledge (``vehicles'') can accelerate convergence and improve accuracy.  
Results in Fig.~\ref{fig:prior_knowledge} further show that, in general, finer-grained knowledge provides more prominent benefits, proving \name's effectiveness in integrating implicit knowledge from pre-trained neural models. 

\begin{figure}[t!]
\centering
\includegraphics[width=0.45\textwidth]{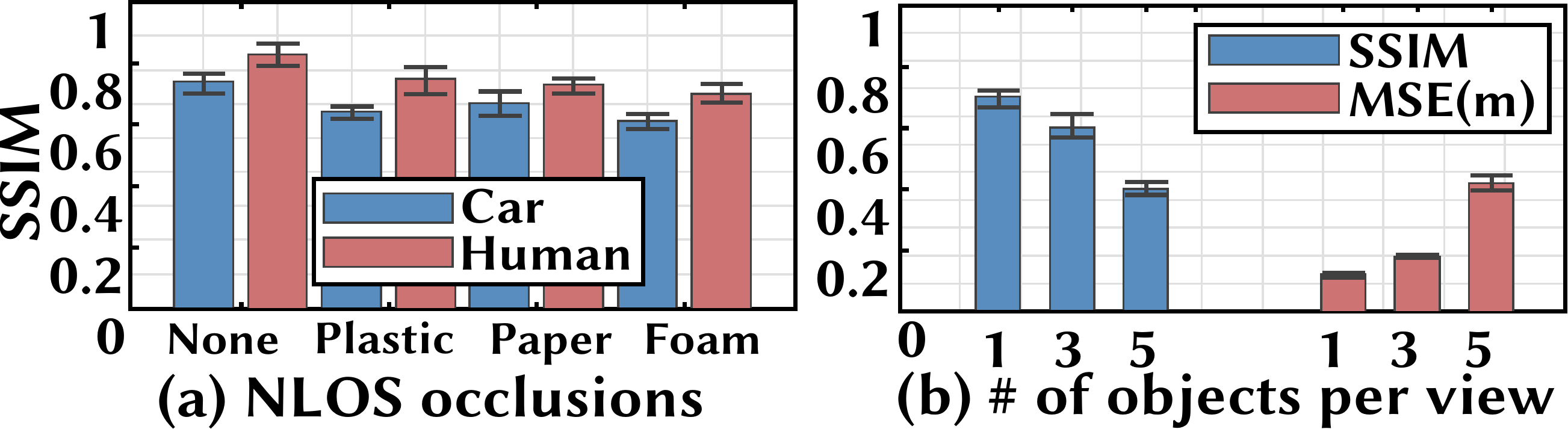}
\caption{(a) 3D reconstruction accuracy under different NLOS occlusions; (b) Impact of the number of objects per radar view.}
\label{fig:result_nlos_n_object}
\end{figure}

\noindent\textbf{Reconstruction in NLOS scenarios.}
We evaluate \name under non-line-of-sight (NLOS) conditions using three types of obstacles: a plastic board (0.7~cm thick), a paper cardboard (0.3~cm thick), and a foam board (2~cm thick). The radar signals reflected by target objects (cars and humans with different postures) are further attenuated and scattered by the obstacles. The results are illustrated in Fig.~\ref{fig:result_nlos_n_object}(a). Even \textit{without prior knowledge} about the occlusion and the objects, the reconstruction performance still preserves the rough shape and pose (SSIM above 0.74 for posture and 0.75 for object shape reconstruction) due to the dominant LoS paths.

Notably, \name achieves reconstruction performance comparable to or even exceeding state-of-the-art data-driven approaches (\eg, 80-90$\%$ shape similarity \cite{liu2019wireless,xue2021mmmesh,wang2018towards}, 30cm depth error and 30-50cm size error \cite{guan2020through}), despite those works relying on larger apertures (\eg, SAR) or shorter sensing distances. In contrast, \name is not restricted to specific radar hardware or object categories and requires no training data. Besides, \name can be combined with data-driven models to enhance their generalization (Sec.~\ref{sec:usecase1}).

\noindent\textbf{Multi-object scenario.} 
\name can reconstruct complex multi-object scenes from a \textit{single} unified sphere initialization, without prior knowledge of the number of objects. We evaluate diverse scenes containing 1-5 objects (\eg, humans, vehicles, and furniture) using the same hardware setup as above. Objects are randomly placed and separated by at least 5 degrees within the radar's field of view. As shown in Fig.~\ref{fig:result_nlos_n_object}(b), \name achieves high fidelity on single-object scenes (SSIM $>$0.8, MSE $\approx 0.2$ m), while performance gradually degrades with more objects (SSIM drops to $\sim$0.5, MSE rises to $\sim$0.4 m for 5 objects). This trend reflects stronger multipath interference and overlapping echoes in crowded scenes, but the reconstruction quality remains acceptable.

We further conduct a stress test to examine the scalability of \name. With a single static radar view, the radar’s spatial resolution becomes the bottleneck beyond 6 objects, occasionally missing smaller items. In contrast, 
when jointly optimizing from multiple radar views (6 radar locations in a surrounding configuration), we observe no noticeable degradation even with more than 10 objects. Multi-view observations provide complementary constraints that mitigate multipath ambiguity and overlapping reflections, enabling \name to scale to dense multi-object environments. The remaining experiments focus on the effectiveness of \name’s differentiation methods and the new capabilities it enables.

\begin{figure}
\centering
\includegraphics[width=0.45\textwidth]{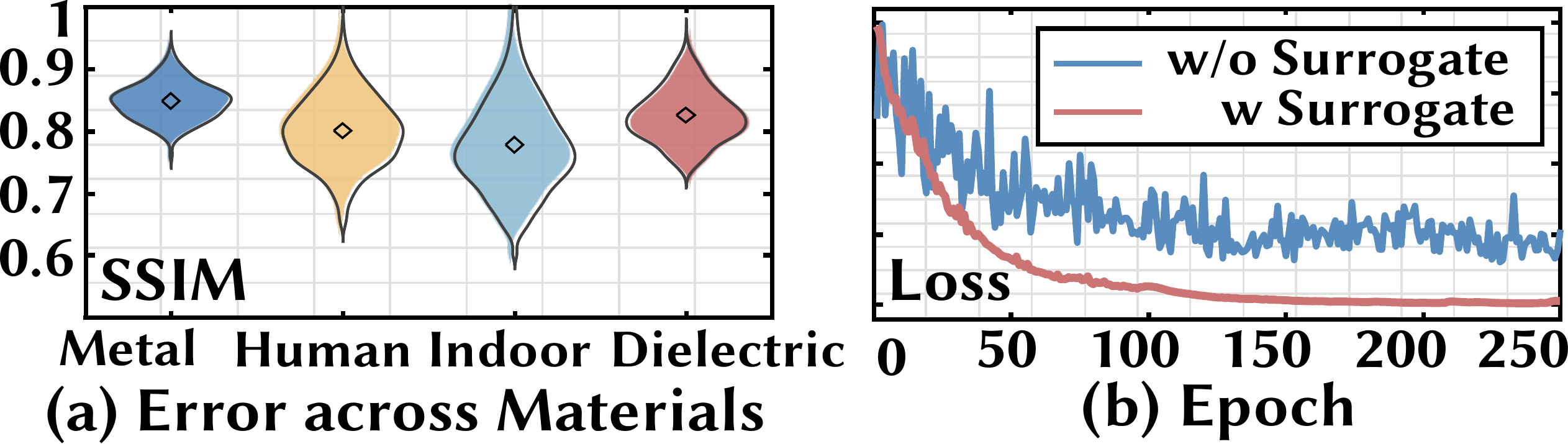}
\caption{(a) Impact of Material for DT Reconstruction; (b) Effectiveness of the Surrogate model.}
\label{fig:results_material_and_surroagate}
\end{figure}

\noindent\textbf{Impact of different materials.}
Fig.~\ref{fig:results_material_and_surroagate}(a) reports the reconstruction error across
representative material categories, including metals, humans, indoor objects,
and dielectrics. While all categories achieve high similarity (SSIM$>0.75$),
metallic surfaces reach the best accuracy due to their more deterministic
reflection, whereas human bodies and cluttered indoor environments are slightly
more challenging. Notably, the performance variation across categories is within
0.1, indicating that \name generalizes well across diverse materials without
requiring material-specific tuning.

\noindent\textbf{Effectiveness of surrogate model.}
Fig.~\ref{fig:results_material_and_surroagate}(b) shows the loss when 
reconstructing a human subject. Without our surrogate for signal-domain
transform (Sec.~\ref{sec:surrogate}), the optimization quickly falls into
oscillations and takes 200+ epochs to stabilize. In contrast, with
the surrogate, the loss consistently decreases and converges within the first 50
epochs. This indicates the surrogate significantly accelerates
convergence ($\approx 4 \times$ faster) and improves stability, which is crucial
for scaling the system to larger and more complex scenes.

\begin{figure}[t!]
\centering
\includegraphics[width=0.49\textwidth]{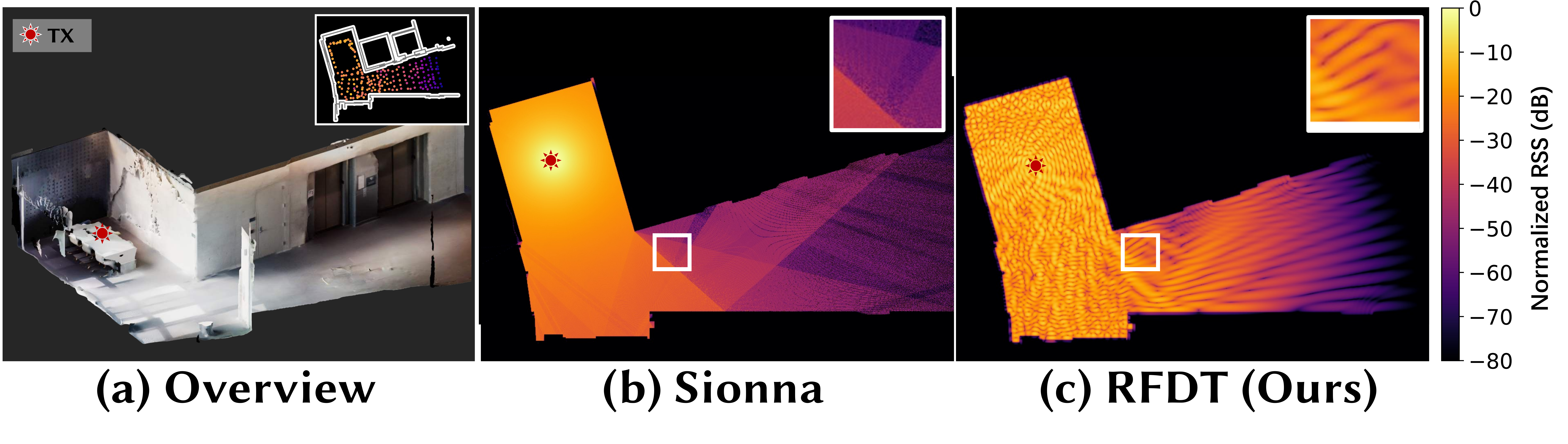}
\caption{\revisetwo{Normalized Received Signal Strength (RSS) in a given scene. (a) The 3D-scanned environment and floor layout, with actual measurement points shown in the top-right inset. \revisethree{(b) Radio map simulated by Sionna (non-coherent). (c) Results generated by \name (coherent).} }
}
\label{fig:rm_compare}
\end{figure} 

\revisetwo{
\noindent\textbf{Performance in complex environments.}
We evaluate \name against Sionna for radio map estimation within a complex indoor environment characterized by diverse materials, including concrete walls, metal chairs, and wooden tables. Ground truth RSS measurements were collected using an ASUS RT-AC86U router (802.11ac) serving as the Access Point (AP) and an iPhone 14 Pro as the receiver client. The AP operated at 2.4 GHz. To ensure consistent spatial sampling, we employed a TurtleBot~\cite{turtlebot4-usermanual} to traverse the environment, collecting approximately 300 samples across a 700\,$\text{ft}^2$ area (approximately $0.428$ samples per $\text{ft}^2$). The environment's 3D geometry was reconstructed using PolyCam \cite{polycam-website}, with material properties manually assigned based on the ITU-R P.2040 database~\cite{ITU-R-P2040}. The environment geometry and radio maps are shown in Fig.~\ref{fig:rm_compare}.

We evaluate the performance of both simulators against the ground truth RSS at the exact same measurement points.
The evaluation metrics include the median error~\cite{ma2024automs}, the Root Mean Square Error (RMSE), and the error standard deviation (SD). \name achieves a median error of $5.2\,\text{dB}$ and an RMSE of $8.46\,\text{dB}$ (SD: $6.67\,\text{dB}$), whereas Sionna yields a larger median error of $7.7\,\text{dB}$ and a substantially higher RMSE of $14.28\,\text{dB}$ (SD: $12.03\,\text{dB}$). We observe that Sionna suffers from large error spikes, particularly at shadow boundaries and within deep shadow regions (as visualized in Fig.~\ref{fig:rm_compare}).
\revisethree{These errors arise because Sionna's non-coherent summation ignores the phase interference required for physical continuity at UTD boundaries, leading to artificial discontinuities. In contrast, \name models phase coherence, producing smooth RSS transitions and demonstrating robustness across these complex boundaries.}

In terms of computational efficiency, \name completes the forward simulation in $0.739\,\text{s}$. In comparison, Sionna v0.19 requires $3.235\,\text{s}$, while the optimized Sionna v1.10 requires $1.22\,\text{s}$. Regarding memory overhead, precise apples-to-apples profiling is challenging due to the heterogeneity of the underlying frameworks (TensorFlow, Dr.Jit, PyTorch, and native CUDA). However, by monitoring total GPU memory usage across 10 distinct runs, we observe that \name exhibits memory consumption comparable to Sionna v1.10, whereas Sionna v0.19 consumes approximately double the GPU resources. Notably, \name achieves this efficiency while performing full phase-coherent computations (using \texttt{Complex64} arithmetic), whereas Sionna primarily uses  \texttt{float32}.
}


\subsection{Computational Cost}
\label{sec:computational_cost}

The computation overhead of \name depends on the scene complexity, parameter count, and computational graph size used in
backpropagation. Our \name forward simulation implementation
(Sec.~\ref{sec:implementation}) is efficient, taking approximately 0.01
second per frame on an NVIDIA A6000 GPU even in the aforementioned multi-object
scene.
The main computational bottleneck arises during backpropagation, primarily due
to automatic differentiation (AD). During this process, the compiler must track
all variables and operations in the forward pass and then execute a
corresponding reverse operation ($\mathcal{O}(P \cdot M \log N)$ with $P$
parameters, $M$ rays and $N$ triangles). This AD step takes between 0.05 to 0.2
seconds, depending on parameter count. Another significant
computation involves updating shapes, which takes about 0.03 seconds
($\mathcal{O}(N_v)$ with $N_v$ mesh vertices) and may require additional
validation passes to ensure physical correctness and prevent corrupted meshes.
The total number of iterations required depends on task. For
example, human pose estimation typically requires around 50 iterations, whereas
3D reconstruction may need around 300 iterations due to the larger number of DT
scene parameters. With the increasing number of ray tracing cores (RT Cores) in
modern GPUs and more efficient AD compilers, we anticipate further improvement.

Fig.~\ref{fig:usecase1_result}(a) empirically validates the asymptotic scaling with $M$ rays and $N$ triangles. 
As $M$ increases, more rays are required for the error to converge, yet the overall trend remains sub-linear in $N$. 
For example, the error converges with about 5k rays for a 10k-triangle scene,
10k rays for 50k triangles, and 15k rays for 100k triangles. In practice, a
typical single-object scene (\eg, a car or human body) contains on the order of
$10^4$ to $5\times10^4$ triangles, while multi-object indoor scenes can reach
$10^5$ to $10^6$ triangles. This confirms the theoretical complexity and
illustrates the tradeoff between scene complexity ($N$) and ray sampling budget
($M$).

\section{Case Studies}
\label{sec:case}
\name is a powerful tool that not only addresses open challenges in RF-based DT reconstruction but also enables new capabilities for both data-driven and model-driven methods. We demonstrate two novel use cases for wireless sensing and communication through one \ttf{unified} framework.

\subsection{Use Case I: Test Time Adaptation for Data-Driven RF Sensing}
\label{sec:usecase1}

\noindent\textbf{Background and Motivation.}
The generalization of machine learning models for RF sensing is a long-lasting concern due to the sensitivity of RF signals to multipath environment.  Test time adaptation (TTA) is a method that boosts the model generalization \cite{liang2024comprehensive}. 
For example, source-free domain adaptation~\cite{chidlovskii2016domain} can assign a pseudo labels to unlabeled samples, enabling TTA in unseen test cases.  
However, the pseudo-labeling still relies on knowledge from training data, and falls short when encountering domain shifts.

\noindent\textbf{\name for Enhancing Generalization.}
\name can be applied to adapt the existing sensing model to unseen test cases,
thus improving generalization. Although the target scenes may differ
significantly from the training sites, the underlying physical laws governing RF
signals remain consistent. \name can embed this physical knowledge in its
simulation and optimization to effectively guide the TTA. As an example, we
present a TTA application in human posture recognition following the PINN model
architecture in Fig.~\ref{fig:pinn}. Given unlabeled radar signals reflected
upon a human subject, an existing DNN model outputs a set of physical parameters
describing the posture. These parameters are used to generate a mesh DT
representation of the subject, which is then fed into \name to simulate the
corresponding RF signals. We calculate the loss as the error between the
simulated RF and the original input RF signal, allowing the gradient to
backpropagate and update the model parameters until convergence. 

\begin{figure}[t!]
\centering
\includegraphics[width=0.44\textwidth]{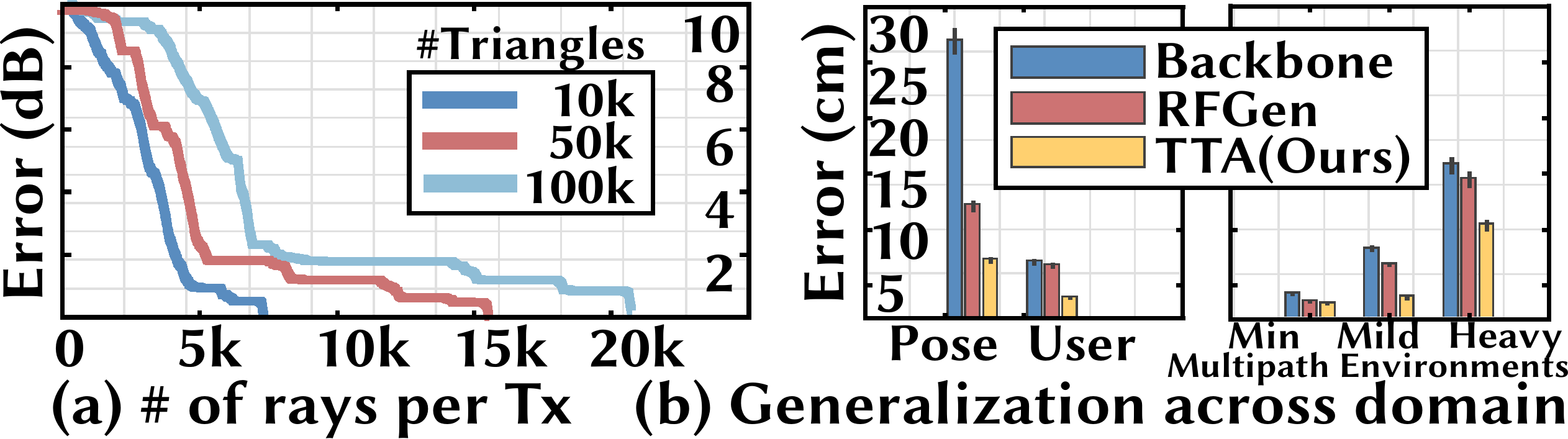}
\caption{ (a) Impact of Ray Density under Scene Complexity; (b) Use case I: Enhancing generalization of ML-based sensing model.}
\label{fig:usecase1_result}
\end{figure}
\noindent\textbf{Experimental Setup and Results.}
We adopt a state-of-the-art DNN-based human posture recognition model, mmMesh
\cite{xue2021mmmesh}, and train it using basic forward-facing postures (T-pose,
A-pose, surrender pose) captured in an outdoor environment with a 
male participant of average build (180~cm height). We implement the TTA model
following Fig.~\ref{fig:pinn}, refining the backbone DNN (\ie, mmMesh~\cite{xue2021mmmesh}) 
with a single, unannotated test sample. As an additional baseline, we
further apply RFGen \cite{chen2023rfgenesis}, an open-source generative
model-based data synthesis method to diversify mmMesh's training data.

We evaluate \name and the baselines on a different subject (female, 160cm height) and a different set of postures (side-views of bending, walking, \etc) than those in the training data, following the setups in RFGen~\cite{chen2023rfgenesis}. The evaluation metric is the vertex error, computed as the average point-wise Euclidean distance (in cm) between the reconstructed mesh and the ground-truth depth-based mesh. The results (Fig.\ref{fig:usecase1_result}(b), left) show that \name reduces the reconstruction error by 75\% and 53\% over mmMesh and RFGen, respectively, on unseen postures. It achieves a similar reduction of 71\% and 65\% on the unseen subject. This improvement stems from its ability to leverage on-site test samples for TTA, overcoming the limited adaptability of generative data augmentation.

\noindent\textbf{Impact of Multipath on Generalization.}
We further investigate the impacts of site-specific multipath effects by running the pose reconstruction in three settings: outdoor open space (minimal multipath), indoor hallway (mild multipath), and indoor office (heavy multipath). The results (Fig.~\ref{fig:usecase1_result}(b), right) show that \name consistently outperforms mmMesh and RFGen, reducing their errors by 38\% to 77\%, respectively. Although RFGen employs an environment diffusion model to diversify the synthetic signals across many visual scenes, it is infeasible to cover every test site. The \name-enabled TTA allows site-specific adaptation of the backbone DNN, thus boosting its robustness and generalization.

\begin{figure}[t!]
\centering
\includegraphics[width=0.45\textwidth]{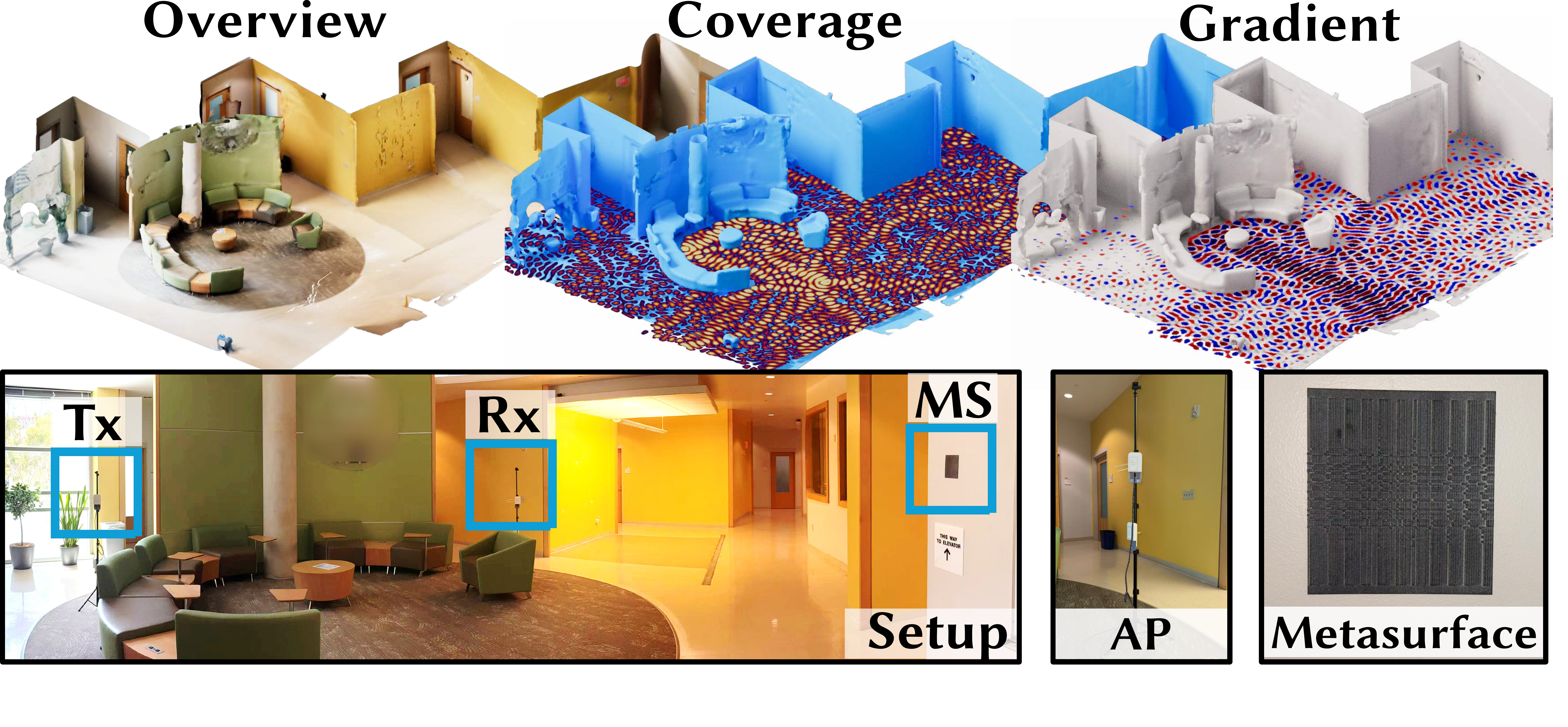}
\caption{Optimizing metasurface (MS) deployment to maximize coverage. Overview: 3D scanned mesh; Coverage: simulated signal strength at each location; Gradient: variation of signal strength \wrt MS translation.}
\label{fig:ms_opt}
\end{figure}

\subsection{Use Case II: Physics-based Parameter Searching for Communication Systems}
\label{sec:usecase3}
\noindent\textbf{Background and Motivation.}
Emerging wireless applications increasingly require searching for optimal parameter configurations in physically constrained systems.
Due to the sensitivity of these systems to physical laws, conventional data-driven models often fall short \cite{yang2020learning,hwang2022solving,raissi2019physics}. 
Current approaches either involve manually crafting symbolic algorithms, which
are difficult to scale, or relying on gradient-free search using complex
simulators, which can be highly time-consuming. We demonstrate how \name can
help overcome such limitations, using metasurface (MS) deployment optimization
as an example. 

MS design
traditionally focuses on optimizing reflection beam patterns
\cite{li2024rfmagus, qian2022millimirror}. Recently, AutoMS \cite{ma2024automs}
introduced a simulation-based method to optimize both reflection patterns and
deployment locations. However, limited by its non-differentiable simulator,
AutoMS relies on exhaustive gradient-free search over all possible 
locations, resulting in a prohibitive computational load. Sionna~\cite{sionna} is differentiable \wrt the MS reflection patterns, but not \wrt
orientations (Sec.~\ref{sec:eval:diff}). 
In contrast, \name can accurately estimate gradients \wrt the MS translation and
deformation parameters, thus enabling gradient-based search for optimal MS
deployment. 

\noindent\textbf{Experimental Setup.}
Our experiments follow a similar setup as in AutoMS \cite{ma2024automs}, where a 60~GHz 802.11ad access point (AP) is placed at a known location in a lab space (Fig.~\ref{fig:ms_opt}). The goal is to optimize the location of a pre-designed MS on the surrounding flat walls, in order to expand the AP's coverage towards a target region. 
The MS is designed to achieve a fixed reflection angle of $45^\circ$ and $15^\circ$ for azimuth and elevation, respectively. The fabrication follows the 3D printing procedure in \cite{woodford2023metasight} (10k unit-cells, $13\times13$ cm$^2$). 

The experiments were conducted in a hallway ($\sim$50 m$^2$) with drywall, wood,
and mixed metallic/non-metallic furniture. We scanned the environment using a 
camera, and used the resulting 3D mesh as input for \name. The optimization
search space allows MS translation along the wall within $\pm3$ m and rotation
within $\pm30^\circ$.

\begin{figure}[t!]
\centering
\includegraphics[width=0.48\textwidth]{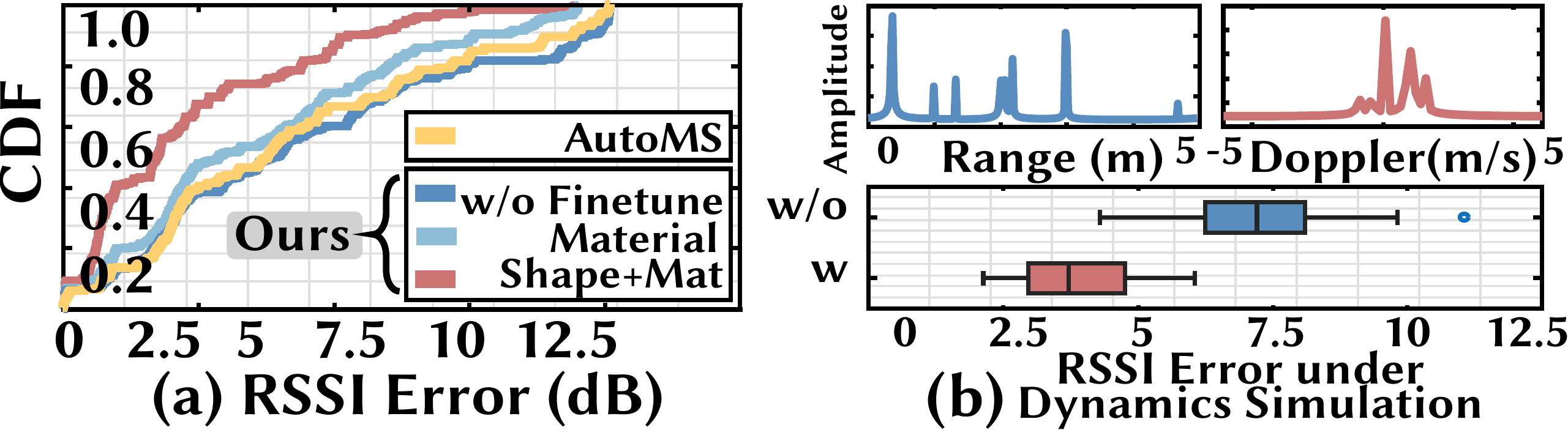}
\caption{(a) RF field simulation accuracy, w and w/o optimizing shape/material parameters in the digital twin; (b) Impact of scene dynamics simulation. }
\label{fig:usecase2}
\end{figure}

\noindent\textbf{Channel Simulation Accuracy.}  
We first evaluate the forward simulation accuracy by comparing the simulated RSS of \name with actual measured samples. A total of 20 samples are collected, uniformly distributed across locations within the scene.  
We compare against AutoMS \cite{ma2024automs}, which also supports tuning ray tracing with real samples but restricts adaptation to material parameters only. 
As shown in Fig.~\ref{fig:usecase2}(a), AutoMS yields a median RSS error of 2.9\,dB. In contrast, our approach jointly adapts both geometry and material, reducing median error to 1.8\,dB, thus narrowing the sim-to-real gap.

\noindent\textbf{Simulation of Scene Dynamics.}  
We further evaluate \name in dynamic scenarios with moving subjects, including
human bodies, chairs, doors, \etc, whose material properties
are obtained from \cite{ITU-R-P2040}. 
Fig.~\ref{fig:usecase2}(b) illustrates the signal spectrum and error statistics. 
The top panels show representative range and Doppler profiles generated by our dynamic simulator, capturing multipath delays and velocity-induced frequency shifts. 
The bottom boxplot compares the RSS error distribution with and without explicit Doppler simulation in a hallway setting. Without modeling dynamics, the error can exceed 8\,dB, reflecting the difficulty of time-varying multipath. With our Doppler and scene dynamics simulation (Sec.~\ref{sec:implementation}), the median error drops to around 3\,dB, demonstrating that \name maintains high fidelity even under scene dynamics.

\noindent\textbf{Metasurface Optimization for Signal Coverage.}
AutoMS relies on gradient-free search and takes about 8 hours on a
consumer-grade PC with an A6000 GPU, whereas \name leverages gradient descent
and finds an optimal location within only 15 minutes on the same machine. This
optimized deployment further improves the RSS at the target region (Rx in
Fig.~\ref{fig:ms_opt}) by 5.7~dB for static scene and by 5.1~dB under random small environmental dynamics.

\section{Conclusion}

We have introduced the design of \name, the first fully differentiable simulation system that solves the inverse problem of RF simulation in order to reconstruct digital twins using RF signals. By efficiently estimating the gradients of simulated signals \wrt various scene parameters, \name uniquely deduces simulation inputs from outputs. 
Through demonstrations across diverse use cases within one unified framework, we envision \name as a pioneering step, opening new research directions in computational RF for future studies.

\begin{acks}

We appreciate the insightful comments and feedback from
the anonymous reviewers and shepherd. This work is partially supported by the
NSF under Grants 2127544, 2403124, 2408393, 2346550, 2312715, 2238839, 2426470, and 2436428, and by an NVIDIA Academic Grant. 
\end{acks}


\newpage
\bibliographystyle{plain}
\bibliography{sigproc}

\appendix

\vspace{-5pt}
\revise{
\section{Limitations and Discussion}
\label{sec:limitation}
\noindent\textbf{Ray Tracing Approximation.} \name achieves full differentiability for ray-tracing-based RF simulation; however, it operates under the high-frequency and far-field assumptions inherent to geometric optics. Extending differentiability to near-field propagation or full-wave electromagnetic solvers remains an important direction for future work.

\noindent\textbf{Gradient Variance and Conditioning.} \name employs symbolic surrogates to mitigate the local non-convexity arising from phase periodicity. However, the underlying landscape often remains \textit{ill-conditioned}, inevitably leading to high gradient variance in complex optimization tasks. Future studies could extend the surrogate framework with data-driven statistical methods, such as adaptive preconditioners or learned gradient estimators, to further address this instability.

\noindent\textbf{Higher-Order Interactions.} \name requires explicit diffraction tracing to model higher-order diffraction and multi-bounce reflections, which can incur computational overhead as the interaction order increases. In practice, this cost can be mitigated by truncating the simulation to first- and second-order diffractions. While this approximation introduces modeling bias, it remains sufficient for most practical use cases where higher-order contributions are negligible.
}

\revisethree{ \noindent\textbf{Coherent vs. Non-Coherent RT.} Unlike typical coverage map tools that use non-coherent summation to approximate the average large-scale path loss~\cite{jaensch2024radiomapestimation}, \name enforces coherent summation, usually reserved for sensing~\cite{10643616}. This is essential for \textit{differentiability}: mathematical continuity at shadow boundaries—required for UTD gradient computation—is preserved only through the interference of coherent components. }

\revise{
\noindent \textbf{Optional Prior Knowledge.}
\label{sec:knowledge}
A key advantage of \name is that it can incorporate cross-domain priors to tame the large, high-dimensional optimization problem in inverse RF rendering. Instead of directly optimizing unconstrained geometry/material parameters (which is expensive and highly non-convex), we restrict the learnable parameters $\Theta$ to a low-dimensional \emph{latent manifold} induced by pretrained, structured shape models. This reduces degrees of freedom, improves gradient conditioning, and typically makes optimization more stable under sparse RF observations.

For human-centric scenes (e.g., Fig.~\ref{fig:gradient_flow}), we adopt the SMPL body model \cite{loper2023smpl}, which represents a mesh with 6,890 vertices using only $10$ (shape) $+$ $72$ (pose) $+$ $3$ (global translation) parameters. \name optimizes these compact parameters and decodes them into a full-resolution mesh that is then passed to the differentiable RF ray-tracing engine. Similarly, in automotive perception, the DT can be constrained to ``roadway objects'' by leveraging pretrained 3D shape decoders from ShapeNet \cite{shapenet2015}: we represent unknown shapes as dynamic density fields at a target resolution (e.g., $512^3$ samples), compress them into compact latent codes, and decode them back into mesh representations for simulation. Unlike purely data-driven RF sensing, this decouples graphics priors from RF simulation and reduces the need for laborious RF training data collection. While stronger priors shrink the search space and ease optimization, they can reduce flexibility for out-of-distribution targets, reflecting a bias--variance trade-off.
}

\noindent\textbf{Bias-Variance Tradeoff.}  
The optimizable parameters in \name are task-dependent, and leveraging prior knowledge (Sec.~\ref{sec:background}) effectively regularizes the parameter space. While extensive observations and strong priors generally enhance optimization outcomes, a fundamental tradeoff exists. Specifically, finer-grained constraints yield higher task accuracy but may hinder generalization, whereas coarser constraints improve robustness at the expense of precision. This bias-variance tradeoff is fundamental to inverse problems across sensing and machine learning.

\section{Future Works}

\subsection{Hessian-based Second Order Optimization}
Second-order optimization via the Hessian matrix enables faster convergence, improved stability, and better handling of ill-conditioned problems. The Hessian metrics provide additional curvature information of the optimization landscape, which allows for more effective escape from saddle points and increased tolerance to hyperparameter variations. The update rule for a damped Newton's method with a general quadratic regularization term is given by:
\begin{equation}
\footnotesize
    \mathbf{x} \leftarrow \mathbf{x} - \eta \left( \frac{\partial^2 \Phi}{\partial \mathbf{x}^2} + \lambda \mathbf{L} \right)^{-1} \left( \frac{\partial \Phi}{\partial \mathbf{x}} + \lambda \mathbf{L} \mathbf{x} \right)
\end{equation}
When the objective function $\Phi$ is a composite function, such as $\Phi(R(\mathbf{x}))$, its second derivative requires the chain rule. For a single variable case, where $\Phi = \Phi(R(x))$, the correct chain rule for the second derivative is:
\begin{equation}
\small
    \frac{d^2\Phi}{dx^2} = \frac{d^2\Phi}{dR^2} \left(\frac{dR}{dx}\right)^2 + \frac{d\Phi}{dR} \frac{d^2R}{dx^2}
\end{equation}

\begin{figure*}[ht!]
        \centering
        \setlength{\abovecaptionskip}{0pt}
        \includegraphics[width=1\linewidth]{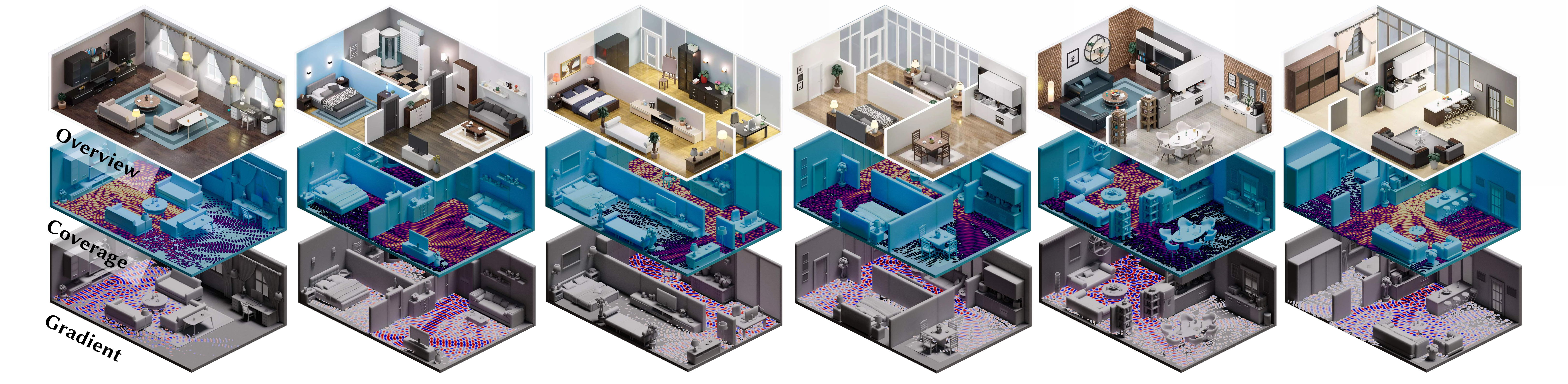}
        \caption{\name estimates the gradient of the coverage map \wrt Tx translations $\theta$ for optimization.}
        \label{fig:rooms}
\vspace{-9pt}
\end{figure*}

\section{Additional Experiments and Details}

\revise{This section provides comprehensive details regarding the experimental setups and extended evaluation results presented in Section~\ref{sec:eval}. We describe the specific hardware configurations for real-world data collection, expanded simulation environments, and comparisons against full-wave solvers and baseline methods to further validate the accuracy and robustness of \name}

\subsection{Real World Setup and Data Collection}
\label{appendix:setup}

We collect the raw radar signals using Vayyar TrigB \cite{vayyar_2022} and TI-AWR1843 radar \cite{AWR1843BOOST} along with an Intel RealSense depth camera. The radar is 1.2~m above the ground so that its field-of-view covers the targets. The measured objects remain stationary during the sensing process. For each target object/scene, we collect 20 seconds of radar samples. For the Vayyar TrigB radar, we use its 77~GHz step-frequency waveform and collect 20 frames, which are averaged to reduce noise. For the TI-AWR1843 radar, we follow the configuration of mmMesh: the radar operates at 77~GHz, with each chirp starting 6~$\mu\text{s}$ after trigger and sweeping linearly at a slope of 60~\text{MHz}/$\mu\text{s}$. Each chirp collects 256 ADC samples at a rate of 4.4~Msps with 30~dB receiver gain, lasts about 65~$\mu\text{s}$ with a 7~$\mu\text{s}$ idle time between chirps, and is repeated in 128 chirp loops per frame with a periodicity of 100~ms. We use one transmission profile with three chirps per loop, and enable automatic start.

 \subsection{Evaluation on Large Simulated Indoor Environment}
We evaluate \name in a large-scale simulated indoor environment containing complex room layouts, diverse geometric structures, and heterogeneous materials (Fig~\ref{fig:rooms}). We consider two representative carrier frequencies, 5~GHz and 60~GHz, corresponding to typical WiFi and mmWave bands. Our simulator stably computes the coherent superposition of multipath components and provides accurate gradients of the received signal field with respect to the transmitter position. These results demonstrate both the scalability and numerical stability of \name when applied to challenging indoor propagation scenarios.

\begin{figure}[t!]
\centering
\includegraphics[width=0.365\textwidth]{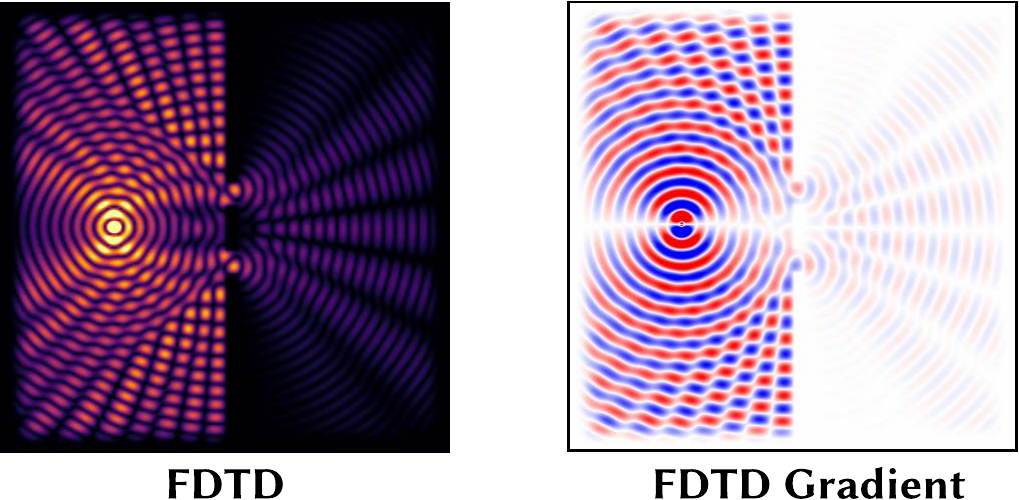}
\vspace{-4pt}
\caption{Finite-Difference Time-Domain Full-wave Simulation Results. Left: RF Field Simulation; Right: Gradient \wrt Tx translation.}
\label{fig:fdtd_result}
\end{figure}

\subsection{Full-wave Simulation}
Although \name primarily focuses on the differentiability rather than accuracy of ray tracing itself, we have conducted experiments to verify the accuracy of its forward ray tracing simulation, 
We compare against finite-difference time-domain (FDTD) solvers~\cite{kunz1993finite}, which compute full-wave solutions to Maxwell's equations. FDTD provides a highly reliable reference, though differences may arise from scene representation (Yee grids vs. triangle meshes, leading to different object thickness), source modeling (voxel emitter in FDTD vs. point source in ray tracing), numerical dispersion, and boundary conditions.  

In our setup, we use 15 layers of perfectly matched layers (PML) to emulate open-space boundary conditions, and discretize the domain with a grid resolution of $\lambda/20$ per cell. Fig.~\ref{fig:fdtd_result} shows the steady-state fields obtained from the FDTD simulation. To enable gradient-based comparison, we further apply the same finite-difference method as in Sec.~\ref{sec:eval:diff}, computing the derivative of received field strength \wrt the transmitter’s $y$-axis displacement.  

Despite the intrinsic differences in discretization and boundary modeling, our ray tracing engine captures all direct, reflected, and diffracted components, which baseline methods such as Sionna fail to model, and \name's gradient trends closely match those of the FDTD solver, demonstrating physical correctness.

\subsection{Sionna Configurations}
\label{appendix:sionna}

We adopt Sionna \cite{sionna} as one of our baselines, an open-source RF simulation library developed by NVIDIA. 
Sionna provides differentiability with respect to limited parameters such as material properties, 
but with no support of differentiation with respect to geometric scene parameters. 
At the time of submission, the latest release was version~v1.10 (June 2025) \cite{aoudia2025sionna}, 
a major refactor of the earlier v0.19 (Sept 2024). 
Notably, the diffraction module available in v0.19 was removed in v1.10 \cite{sionna_discussion_838}. 
However, both versions remain non-differentiable with respect to geometry: 
neither the \texttt{coverage\_map} operator in v0.19 nor the \texttt{radiomap} operator in v1.10 
propagates gradients from positional parameters to the output coverage maps.

\begin{figure}[t!]
\centering
\includegraphics[width=0.4\textwidth]{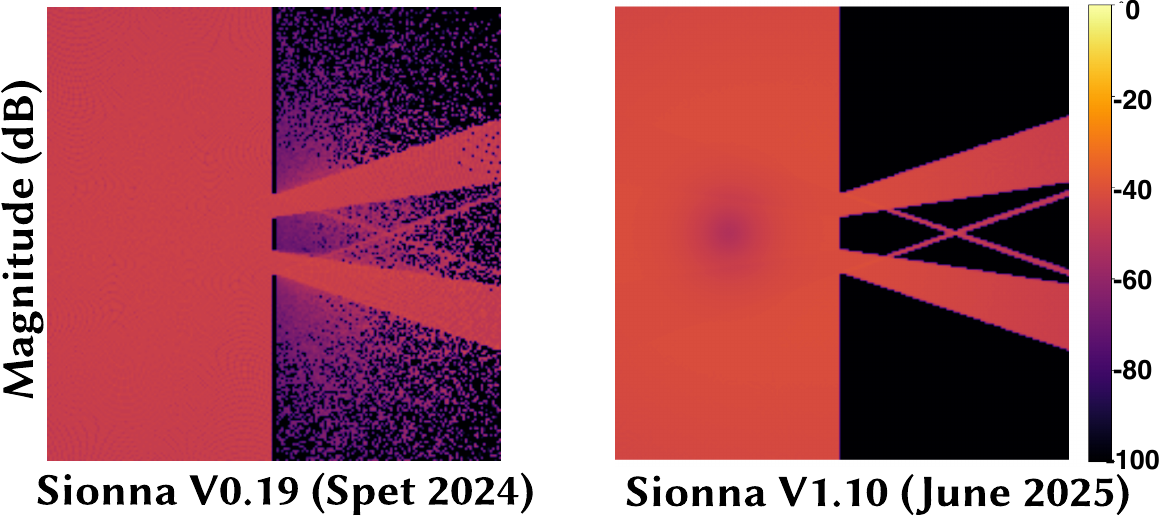}
\caption{Sionna results under different versions. Version 0.19 includes diffraction simulation, whereas version 1.10 has removed this feature. Both results significantly differ from the full-wave solution due to non-coherent combinations.}
\vspace{-4pt}
\label{fig:sionna_results}
\end{figure}

\revisethree{
\noindent\textbf{Solver Architectures: Path vs. Radio Map.}
It is important to distinguish between Sionna's \texttt{path} solver and its \texttt{radiomap} (or \texttt{coverage\_map}) solver. While the \texttt{path} solver supports coherent combination with full phase information, it is designed for sparse, point-to-point link simulations. Adapting it for dense radio map generation is computationally prohibitive; for instance, generating a standard $512 \times 512$ heatmap requires instantiating over 260,000 individual receiver objects. In our tests, this approach consistently caused memory exhaustion and system freezes even on high-end NVIDIA A6000 (48GB) GPUs.
Consequently, for coverage prediction tasks, we utilize the dedicated \texttt{radiomap} solver. However, our inspection reveals that this solver performs simplified \textit{non-coherent} power summation~\cite{nv_sionna_solver_cm}. It discards phase information and aggregates scalar power values, thereby failing to capture constructive or destructive interference patterns. Although v0.19 nominally included a diffraction energy term, the results were inconsistent and physically incorrect (Fig.~\ref{fig:sionna_results}).
}

\revise{
\noindent\textbf{Configurations.}
We use Sionna’s path solver for radar simulation and the coverage map solver for radio field simulation. For all coverage map experiments, we use the same settings for Sionna and \name. The number of samples per transmitter is set to one million, and the maximum depth is set to three. The cell size is set to one-eighth of the wavelength (e.g., approx. 0.016 meter at 2.4 GHz).

\noindent\textbf{Modification for radar simulation.}
Since Sionna by default does not support radar simulation, 
we extended its functionality by simulating separate transmit and receive antennas 
and constructing propagation paths whose phases can be computed from distance. 
Instead of performing ray tracing for every antenna element, we compute propagation paths once from the array centroid and derive the full MIMO response based on the relative phase offsets determined by the array geometry.
This modification ensures compatibility between TensorFlow-based operators and the underlying Mitsuba ray-tracing engine, 
allowing us to obtain differentiable channel impulse responses (CIRs). 

Using these computed paths, we synthesize the radar return by aggregating the contributions of individual rays. For the widely used FMCW radar~\cite{iovescu2020fundamentals}, the received signal is modeled as
\begin{equation}
\small
S_{\mathrm{IFS}}(t) = \sum_{i=0}^{N} A(\alpha,\gamma)\exp\!\left( 2\pi j(\mu t + f_c t) \right),
\label{eq:ifs_signal}
\end{equation}
where $N$ denotes the total number of propagation rays. The term $A(\alpha,\gamma)$ represents the antenna radiation pattern evaluated at azimuth angle $\alpha$ and elevation angle $\gamma$. The carrier frequency is given by $f_c$, while $\mu$ denotes the chirp rate, defined as $\mu = B/T$, with $B$ being the signal bandwidth and $T$ the chirp duration.

However, since this method relies on direct automatic differentiation without handling edge discontinuities, it cannot correctly capture geometric boundaries, resulting in noisy and incorrect gradients.
}

\vspace{-4pt}
\section{Proofs}

\revise{
We provide the formal mathematical proofs and derivations that underpin the differentiable rendering theories introduced in Section~\ref{sec:diffRT}. We rigorously establish the incompatibility of Monte Carlo integration with point-source specular reflection discussed in Section~\ref{sec:discontinuity} and derive the asymptotic properties of the edge diffraction transition function that ensure the smooth gradient propagation utilized in Section~\ref{sec:unbiased_diff}
}

\subsection{Measure-Zero Nature of Specular Paths}
\label{appendix:proof_point_specular}

\textbf{Background and Preliminaries.}
We begin with the standard rendering equation at a surface point with unit normal $n$ and outgoing direction $\omega_o$:
\begin{equation}
\small
L_o(\omega_o) \;=\; \int_{\Omega^+} f_r(\omega_i,\omega_o)\, L_i(\omega_i)\, (n \cdot \omega_i)\, d\omega_i,
\end{equation}
where $\Omega^+$ is the upper hemisphere, $L_i$ is incoming radiance, and $f_r$ is the Bidirectional Reflectance Distribution Function (BRDF).  

We recall two idealized models:
\textit{Perfect Specular BRDF.}  
For an ideal mirror with specular albedo $\rho_s \in [0,1]$, the BRDF enforces the law of reflection using a Dirac delta distribution on the direction domain:
\begin{equation}
\small
f_r(\omega_i,\omega_o) \;=\; \rho_s \,\frac{\delta_{\Omega}(\omega_i - R(\omega_o))}{|n \cdot \omega_i|},
\end{equation}
where $R(\omega_o)$ denotes the unique incident direction that reflects to $\omega_o$. The factor $1/|n \cdot \omega_i|$ guarantees correct energy normalization under the cosine factor.

\textit{Point Source.}  
A point light arriving from direction $\omega_l$ is expressed as:
\begin{equation}
\small
L_i(\omega_i) \;=\; L_{\text{src}} \cdot \delta_{\Omega}(\omega_i - \omega_l).
\end{equation}
We now substitute the specular BRDF and the point source models into the rendering equation to find the outgoing radiance $L_o(\omega_o)$:
\begin{equation}
\footnotesize
L_o(\omega_o) \;=\; \int_{\Omega^+} \left( \rho_s \frac{\delta_{\Omega}(\omega_i - R(\omega_o))}{|n \cdot \omega_i|} \right) \left( L_{\text{src}} \cdot \delta_{\Omega}(\omega_i - \omega_l) \right) (n \cdot \omega_i)\, d\omega_i.
\end{equation}
By rearranging terms, we get:
\begin{equation}
\small
L_o(\omega_o) \;=\; \rho_s L_{\text{src}} \int_{\Omega^+} \frac{(n \cdot \omega_i)}{|n \cdot \omega_i|} \delta_{\Omega}(\omega_i - R(\omega_o))\, \delta_{\Omega}(\omega_i - \omega_l)\, d\omega_i.
\end{equation}
As the integration is performed over the upper hemisphere $\Omega^+$, the condition $(n \cdot \omega_i) \ge 0$ holds, which means $|n \cdot \omega_i| = (n \cdot \omega_i)$. The cosine term therefore cancels out, simplifying the integral to:
\begin{equation}
\small
L_o(\omega_o) \;=\; \rho_s L_{\text{src}} \int_{\Omega^+} \delta_{\Omega}(\omega_i - R(\omega_o))\, \delta_{\Omega}(\omega_i - \omega_l)\, d\omega_i.
\label{eq:appendix_integral_two_deltas}
\end{equation}
To evaluate this integral, we rely on the fundamental \textbf{sifting property} of the Dirac delta distribution. For a general function $g(x)$ and a delta function $\delta(x-a)$, the property is defined as:
\begin{equation}
\small
\int g(x) \, \delta(x - a) \, dx = g(a).
\end{equation}
We apply this property to Equation~\ref{eq:appendix_integral_two_deltas} by treating $g(\omega_i) = \delta_{\Omega}(\omega_i - R(\omega_o))$ as the function to be evaluated, and $\delta_{\Omega}(\omega_i - \omega_l)$ as the sifting delta. The integral is thus evaluated by substituting $\omega_i = \omega_l$ into $g(\omega_i)$:
\begin{equation}
\small
\int_{\Omega^+} \delta_{\Omega}(\omega_i - R(\omega_o))\, \delta_{\Omega}(\omega_i - \omega_l)\, d\omega_i \;=\; \delta_{\Omega}(\omega_l - R(\omega_o)).
\end{equation}
This leads to the final expression for the outgoing radiance:
\begin{equation}
L_o(\omega_o) \;=\; \rho_s L_{\text{src}} \cdot \delta_{\Omega}(\omega_l - R(\omega_o)).
\end{equation}

This result has a clear physical interpretation: the outgoing radiance is zero for all directions except for the one where the view direction $\omega_o$ is the perfect specular reflection of the light source direction $\omega_l$. This occurs when the argument of the delta function is zero, i.e., $\omega_l = R(\omega_o)$, which is equivalent to the law of reflection, $\omega_o = R(\omega_l)$.

\noindent\textbf{Inefficiency of Naïve Monte Carlo Estimator.}
A standard Monte Carlo (MC) estimator samples $K$ directions $\{\omega_{i,k}\}$ from a continuous probability density $p(\omega)$ and computes
\begin{equation}
\footnotesize
\widehat{L}_o = \frac{1}{K} \sum_{k=1}^K 
\frac{f_r(\omega_{i,k},\omega_o)\,L_i(\omega_{i,k})\,(n \cdot \omega_{i,k})}{p(\omega_{i,k})}.
\end{equation}

In our case, the integrand is supported only at the single direction $\omega^* = R(\omega_o) = \omega_l$. Mathematically, a single direction in the continuous angular domain constitutes a set of \textbf{measure zero}. For any absolutely continuous density $p(\omega)$,
\begin{equation}
\Pr[\omega_{i,k} = \omega^*] = 0,
\end{equation}
so with probability one, no sample hits the support of the integrand. Consequently,
\begin{equation}
\widehat{L}_o = 0 \quad \text{almost surely},
\end{equation}
even when the true integral equals $\rho_s L_{\text{src}} > 0$.  

Formally, the product of deltas defines a measure that is singular with respect to $p(\omega) d\omega$, so the Radon--Nikodym derivative required for unbiased importance sampling does not exist. This explains why the Monte Carlo estimator collapses.

\vspace{-4pt}
\subsection{Differentiable Ray Tracing via Edge Diffraction}
\label{appendix:proof_diffraction}

\textbf{Diffraction Point Construction.}
We recall that a wedge is defined by two adjacent planar faces
$\triangle_a,\triangle_b$ sharing an edge $\mathcal{E}$ parameterized as
$\mathbf{e}(t)=\mathbf{v}_0+t\hat{\mathbf{e}}$ with $\hat{\mathbf{e}}=\frac{\mathbf{v}_1-\mathbf{v}_0}{\|\mathbf{v}_1-\mathbf{v}_0\|}$.
Given transmitter and receiver locations $\mathbf{x}_{\mathrm{tx}},\mathbf{x}_{\mathrm{rx}}$, the optical path length via a point $\mathbf{e}(t)$ is
\begin{equation}
d(t) = \|\mathbf{x}_{\mathrm{tx}}-\mathbf{e}(t)\|+\|\mathbf{e}(t)-\mathbf{x}_{\mathrm{rx}}\|.
\end{equation}
By Fermat’s principle, the diffraction point $\mathbf{x}_d=\mathbf{e}(t_d)$ is defined as the unique minimizer of $d(t)$ over $t\in[0,L]$.  
Convexity of each distance term implies strict convexity of $d(t)$ except for measure-zero degenerate cases, ensuring existence and uniqueness of $t_d$.

\medskip
\noindent
\textbf{Connection to Specular Reflection.}
For a single planar face $\Pi$ with unit normal $\mathbf{n}$, the specular point $\mathbf{x}_\Pi$ is obtained by minimizing the same optical length subject to $\mathbf{x}\in\Pi$.  
Stationarity leads to the condition
\begin{equation}
\frac{\mathbf{x}-\mathbf{x}_{\mathrm{tx}}}{\|\mathbf{x}-\mathbf{x}_{\mathrm{tx}}\|}-
\frac{\mathbf{x}_{\mathrm{rx}}-\mathbf{x}}{\|\mathbf{x}_{\mathrm{rx}}-\mathbf{x}\|}\;\;\perp\;\; T_\Pi,
\end{equation}
where $T_\Pi$ denotes the tangent space of the plane.  
This is precisely the law of reflection: tangential components of incidence and exit directions agree, while normal components are mirrored.

\medskip
\noindent
\textbf{Limiting Argument on the Wedge.}
Consider the wedge formed by $\Pi_1,\Pi_2$ with edge direction $\hat{\mathbf{e}}=\mathbf{n}_1\times\mathbf{n}_2$.  
For each $\varepsilon>0$, offset planes $\Pi_i(\varepsilon)$ parallel to $\Pi_i$ admit unique specular points $\mathbf{x}_i(\varepsilon)$.  
As $\varepsilon\to 0^+$, the feasible region shrinks toward the edge $\mathcal{E}$.  
The first-order optimality condition on $\mathbf{x}_i(\varepsilon)$ requires that the derivative of the optical path along $\hat{\mathbf{e}}$ vanishes:
\begin{equation}
\frac{(\mathbf{e}(t)-\mathbf{x}_{\text{tx}})^\mathsf{T}\hat{\mathbf{e}}}{\|\mathbf{e}(t)-\mathbf{x}_{\text{tx}}\|}
+\frac{(\mathbf{e}(t)-\mathbf{x}_{\text{rx}})^\mathsf{T}\hat{\mathbf{e}}}{\|\mathbf{e}(t)-\mathbf{x}_{\text{rx}}\|}=0.
\label{eq:stationarity_edge}
\end{equation}
Equation~\eqref{eq:stationarity_edge} coincides exactly with the stationarity condition defining $t_d$.  
Thus both sequences $\mathbf{x}_1(\varepsilon),\mathbf{x}_2(\varepsilon)$ converge to $\mathbf{x}_d$ as $\varepsilon\to 0$.  
In this limiting sense, the unique diffraction point $\mathbf{x}_d$ is simultaneously the specular point for both faces.

\medskip
\noindent
\textbf{Implication.}
We have established that the UTD diffraction point $\mathbf{x}_d$ coincides with the edge point where specular reflection on the two faces becomes indistinguishable.  
Consequently, the specular contribution transitions continuously into a diffracted contribution at the wedge.  
This property is precisely what enables us to differentiate visibility through edges: the optical path length via $\mathbf{x}_d$ depends smoothly on geometric parameters, and the envelope theorem guarantees that its gradient is obtained by differentiating the path length while holding $t_d$ implicit.  
Hence edge diffraction supplies the physically grounded continuous limit required to make specular visibility differentiable.

\subsection{Asymptotics of the Transition Function}
\label{appendix:asymptotic}

Recall the transition function in Eq ~\eqref{eq:transition}.
We prove its limiting behavior as $x\to\infty$ and $x\to 0^+$.

\noindent\textbf{Closed form via the complementary error function.}
Let $a=\sqrt{x}$. Rotate the integration path by setting $\tau=e^{-j\pi/4}u$.
Then $\tau^2=e^{-j\pi/2}u^2=-j\,u^2$ and $e^{-j\tau^2}=e^{-(-j)u^2}=e^{-u^2}$, 
while $d\tau=e^{-j\pi/4}\,du$. The lower and upper limits map as 
$\tau\in[a,\infty)\mapsto u\in[a e^{j\pi/4},\infty e^{j\pi/4}]$. Since $e^{-u^2}$ is entire, 
Cauchy’s theorem permits integration along the straight ray at angle $\pi/4$, yielding
\begin{equation}
\small
\int_{\sqrt{x}}^{\infty} e^{-j\tau^{2}}\,d\tau
= e^{-j\pi/4}\!\int_{a e^{j\pi/4}}^{\infty e^{j\pi/4}} e^{-u^{2}}\,du
= e^{-j\pi/4}\,\frac{\sqrt{\pi}}{2}\,\mathrm{erfc}\!\left(e^{j\pi/4}a\right).
\label{eq:erfc_repr}
\end{equation}
Substituting \eqref{eq:erfc_repr} into \eqref{eq:transition} gives the convenient closed form
\begin{equation}
\colorbox{gray!10}{$
\mathcal{F}(x)
= j\sqrt{\pi}\,e^{-j\pi/4}\,\sqrt{x}\,e^{jx}\,
\mathrm{erfc}\!\big(e^{j\pi/4}\sqrt{x}\big)
$}
\label{eq:F_erfc}
\end{equation}
with the principal branches $\sqrt{x}\ge 0$ and $\sqrt{j}=e^{j\pi/4}$.

\noindent\textbf{Large-argument expansion ($x\to\infty$)}
Let $z=e^{j\pi/4}\sqrt{x}$ so that $z^2=jx$. For $|\arg z|<3\pi/4$ (satisfied here), the complementary error function admits the Poincaré asymptotic series \cite[\S7.12(ii)]{NIST:DLMF}:
\begin{equation}
\footnotesize
\mathrm{erfc}(z)
\sim \frac{e^{-z^{2}}}{\sqrt{\pi}\,z}\left(1-\frac{1}{2z^{2}}+\frac{3}{(2z^{2})^{2}}-\frac{15}{(2z^{2})^{3}}+\cdots\right),
\quad |z|\to\infty,
\label{eq:erfc_asymp}
\end{equation}
with remainder after truncation at order $M$ bounded by 
$O\!\big(|z|^{-2M-1}\big)$ uniformly in any closed subsector $|\arg z|\le \theta<3\pi/4$.

Insert \eqref{eq:erfc_asymp} into \eqref{eq:F_erfc}. The exponential factors cancel since 
$e^{jx}e^{-z^2}=e^{jx}e^{-jx}=1$, and $j e^{-j\pi/4}/z=j e^{-j\pi/4}/(e^{j\pi/4}\sqrt{x})=j e^{-j\pi/2}/\sqrt{x}=-j^{2}/\sqrt{x}=1/\sqrt{x}$. 
Multiplying by $\sqrt{\pi}\sqrt{x}$ and collecting powers of $z^{-2}=(jx)^{-1}$ yields
\begin{equation}
\small
\colorbox{gray!10}{$
\mathcal{F}(x)
= 1 + \frac{j}{2x} - \frac{3}{4x^{2}} + \frac{15j}{8x^{3}} + O\!\left(\frac{1}{x^{4}}\right),
\quad x\to\infty,
$}
\label{eq:F_large_x}
\end{equation}

hence $\lim_{x\to\infty}\mathcal{F}(x)=1$ with an $O(1/x)$ approach. 
The $O(1/x^{m})$ remainder follows from the $O(|z|^{-2m-1})$ remainder in \eqref{eq:erfc_asymp} since $|z|=\Theta(\sqrt{x})$.

\noindent\textbf{Small-argument expansion ($x\to 0^+$).}
Using the Maclaurin series for $\mathrm{erfc}$ \cite[\S7.12(i)]{NIST:DLMF},
\begin{equation}
\footnotesize
\mathrm{erfc}(z) = 1 - \frac{2}{\sqrt{\pi}}z + \frac{2}{3\sqrt{\pi}}z^{3} + O(z^{5}),
\qquad z\to 0,
\label{eq:erfc_small}
\end{equation}
together with $z=e^{j\pi/4}\sqrt{x}$ and $e^{jx}=1+jx+O(x^{2})$, 
substitution into \eqref{eq:F_erfc} gives
\begin{align}
\footnotesize
\mathcal{F}(x)
&= j\sqrt{\pi}\,e^{-j\pi/4}\sqrt{x}\Big(1+jx+O(x^{2})\Big)\\
&
\times \Big(1-\tfrac{2}{\sqrt{\pi}}e^{j\pi/4}\sqrt{x}+\tfrac{2}{3\sqrt{\pi}}e^{3j\pi/4}x^{3/2}+O(x^{5/2})\Big) \nonumber\\
&= \colorbox{gray!10}{$
\sqrt{\pi}\,e^{j\pi/4}\,\sqrt{x}\;-\;2j\,x\;+\;O\!\big(x^{3/2}\big),
\qquad x\to 0^+.
$}
\label{eq:F_small_x}
\end{align}
In particular, $\lim_{x\to 0^+}\mathcal{F}(x)=0$, and the leading-order scaling is 
$\mathcal{F}(x)\sim \sqrt{\pi}\,e^{j\pi/4}\sqrt{x}$, which suppresses the geometric-optics singularity at the edge.

\vspace{-4pt}
\section{Forward Simulation Details.}
\revise{
This appendix details the algorithmic implementation of the forward ray tracing of \name. We elaborate on the reparameterized method of images for deterministic path construction introduced in Section~\ref{sec:reparam}, the physical modeling of material interactions for non-line-of-sight propagation, and the mechanisms for simulating scene dynamics and Doppler effects.
}
\subsection{Reparameterized Method-of-Images}
\label{appendix:method_of_images}
Specifically, for any triangle in the mesh with vertices $(v_0, v_1, v_2)$, we define the corresponding plane via the mapping $S: \mathbb{R}^{3 \times 3} \to \mathbb{R}^3 \times \mathbb{R}^3$:
\begin{equation}
S(v_0, v_1, v_2) = \left(v_0,\; (v_1 - v_0) \times (v_2 - v_0) \right)
\label{eq:plane}
\end{equation}
This representation encodes both a reference point $p_0$ on the plane and its normal vector $\mathbf{n}$. The mirror image of an arbitrary point $\mathbf{p}$ with respect to plane $S = (\mathbf{p}_0, \mathbf{n})$ is computed as:
\begin{equation}
\mathbf{Mirror}(\mathbf{p}, S) := \mathbf{p} - 2 \frac{ (\mathbf{p} - \mathbf{p}_0)\cdot \mathbf{n} }{ \|\mathbf{n}\|^2 } \mathbf{n},  \quad S = (\mathbf{p}_0, \mathbf{n})
\label{eq:mirror}
\end{equation}
Applying the image method recursively, we compute the virtual image of the RX through successive reflections. Starting from the last reflecting surface and working backwards through the sequence of reflecting surfaces, we construct the chain of image points:
\begin{equation}
\mathbf{p}_\text{rx}^{(0)} := \mathbf{p}_\text{rx}, \quad
\mathbf{p}_\text{rx}^{(i)} := \mathbf{Mirror}(\mathbf{p}_\text{rx}^{(i-1)}, S(\triangle_{m-i+1})),
\label{eq:image_point}
\end{equation}

The actual reflection points are then determined by computing the intersection between the ray from the TX to the final virtual RX image and each intermediate reflecting plane:
\begin{equation}
\mathbf{p}_\text{ref}^{(i)} := \textbf{Intersect}_{\text{plane}}( \mathbf{p}_\text{ref}^{(i-1)} \rightarrow \mathbf{p}_\text{rx}^{(m-i)},\; S(\triangle_i) ),  
\label{eq:reflect_point}
\end{equation}

\noindent where $
\mathbf{p}_\text{ref}^{(0)}:= \mathbf{p}_\text{tx}$.
The complete propagation path is thus constructed as the ordered sequence comprising the TX, all intermediate reflection points, and the RX:
\begin{equation}
\small
\begin{aligned}
\mathcal{P}(\mathbf{p}_\text{tx}, \mathbf{p}_\text{rx} \mid \triangle_1,\dots,\triangle_m)
&:= \{\mathbf{p}_\text{tx}, \mathbf{p}_\text{ref}^{(1)}, \dots, \mathbf{p}_\text{ref}^{(m)}, \mathbf{p}_\text{rx}\} \\
= &\,\, \mathcal{T}_{\mathrm{path}}\!\left(\mathbf{p}_\mathrm{tx},\mathbf{p}_\mathrm{rx},\{\triangle(v_{i,0},v_{i,1},v_{i,2})\}_{i=1}^m\right)
\end{aligned}
\end{equation}
The second line rewrites the explicit point sequence as the output of $\mathcal{T}_{\mathrm{path}}$, which encapsulates the preceding reflection and intersection operations from Eq.~\eqref{eq:plane} to Eq.~\eqref{eq:reflect_point} into a single function.
This $\mathcal{T}_{\mathrm{path}}$ is a \textit{deterministic} function that maps a set of triangle vertices to the complete propagation path. 
The resulting path length $L_\mathcal{P}$, path amplitude $A_\mathcal{P}$, and phase are all continuous and differentiable \wrt every vertex $v_{i,j}$, $\mathbf{p}_\text{tx}$, and $\mathbf{p}_\text{rx}$, 
as the mapping is defined via intersections with the infinite planes supporting the triangles.
However, these points may fall outside the actual triangle boundaries, requiring an additional edge validation step.  This validation is performed using barycentric coordinates \cite{moller1997fast}. For a point $\mathbf{p}$ and triangle with vertices $(v_0, v_1, v_2)$,
\begin{equation}
\small
\begin{bmatrix}
u \\
w
\end{bmatrix}
=
\begin{bmatrix}
e_1 \cdot e_1 & e_1 \cdot e_2 \\
e_1 \cdot e_2 & e_2 \cdot e_2
\end{bmatrix}^{-1}
\begin{bmatrix}
e_1 \cdot t \\
e_2 \cdot t
\end{bmatrix},
\end{equation}
where $e_1 = v_1 - v_0$, $e_2 = v_2 - v_0$, and $t = \mathbf{p} - v_0$. The point lies within the triangle if and only if the barycentric coordinates satisfy the constraints $u \geq 0$, $w \geq 0$, and $u + w \leq 1$.

\subsection{Material and NLOS Simulation}
\label{appendix:material_nlos}
In ray-tracing based wireless propagation simulators, the interaction of electromagnetic (EM) waves with materials is modeled through a combination of reflection, transmission, and absorption phenomena. Penetration modeling is essential for capturing non-line-of-sight (NLOS) propagation in RF channels, since many rays traverse or reflect from walls, glass, or other building materials. Each surface in the scene is associated with a material characterized by its complex permittivity $\epsilon_r = \epsilon' - j\epsilon''$, permeability $\mu_r$, and conductivity $\sigma$. These parameters determine the boundary conditions that govern the Fresnel coefficients.

\textbf{\textit{Reflection and Transmission.}}
When a plane wave impinges on a planar interface between two materials, the incident, reflected, and transmitted fields satisfy the continuity of tangential electric and magnetic fields. For a wave incident from medium 1 to medium 2, with relative permittivities $\epsilon_{r,1}$ and $\epsilon_{r,2}$, the Fresnel reflection coefficient for perpendicular (s) and parallel (p) polarizations is given by:
{\footnotesize
\begin{equation}
\Gamma_\perp(\theta_i) = \frac{\eta_2 \cos\theta_i - \eta_1 \cos\theta_t}{\eta_2 \cos\theta_i + \eta_1 \cos\theta_t}, 
\Gamma_\parallel(\theta_i) = \frac{\eta_1 \cos\theta_i - \eta_2 \cos\theta_t}{\eta_1 \cos\theta_i + \eta_2 \cos\theta_t},
\end{equation}
}
where $\eta_k = \sqrt{\tfrac{j\omega\mu_0}{\sigma_k + j\omega \epsilon_0 \epsilon_{r,k}}}$ is the intrinsic impedance of medium $k$, $\theta_i$ is the angle of incidence, and $\theta_t$ is the transmission angle determined by Snell’s law:
\begin{equation}
n_1 \sin\theta_i = n_2 \sin\theta_t, \quad n_k = \sqrt{\epsilon_{r,k}\mu_{r,k}}.
\end{equation}
\textbf{\textit{Penetration and Absorption.}}
To account for material penetration, the transmitted wave is attenuated according to the propagation constant
\begin{equation}
\scriptsize
\gamma = \alpha + j\beta = j\omega \sqrt{\mu\epsilon \left( 1 - j\frac{\sigma}{\omega \epsilon}\right)},
\end{equation}
where $\alpha$ is the attenuation constant and $\beta$ the phase constant. The field magnitude inside the material decays as
\begin{equation}
\small
|E(z)| = |E_0| e^{-\alpha z},
\end{equation}
where $z$ is the propagation depth.

\textbf{\textit{Implementation in Ray Tracing.}}
During simulation, each ray carries a complex amplitude and polarization state. At each surface interaction, the ray is split into reflected and transmitted components according to $\Gamma_{\perp,\parallel}$ and $T_{\perp,\parallel} = 1 + \Gamma_{\perp,\parallel}$. The simulator accumulates the phase shift $e^{-j\beta d}$ and attenuation $e^{-\alpha d}$ along the ray path of length $d$ through the medium. For heterogeneous materials, piecewise propagation is applied, with multiple segments contributing multiplicatively to the final path gain.

\textbf{\textit{Practical Considerations.}}
In practice, tabulated material parameters (e.g., concrete, glass, metal) are used. Conductors are modeled by setting $\sigma \to \infty$, leading to $|\Gamma| \approx 1$. Lossy dielectrics such as walls are modeled with finite $\epsilon''$ that determines $\alpha$. This allows the simulator to capture multipath effects, shadowing, and penetration loss consistent with empirical channel models.

\textbf{\textit{Differentiation.}}
Material parameters appear in the Fresnel coefficients and propagation constant as continuous and differentiable quantities. Since they only affect the complex amplitude of each ray path through multiplicative factors, the dependence of path gain on $\epsilon_r$, $\mu_r$, and $\sigma$ is smooth. This allows analytical gradients, e.g.,
\begin{equation}
\small
\frac{\partial \Gamma_\perp}{\partial \epsilon_r}, \quad 
\frac{\partial \Gamma_\parallel}{\partial \epsilon_r}, \quad 
\frac{\partial e^{-\alpha d}}{\partial \epsilon_r},
\end{equation}
to be obtained in closed form, facilitating gradient-based optimization and integration with learning frameworks.

\subsection{Doppler and Dynamics Simulation}
\label{appendix:doppler}

We extend the ray tracing formulation to account for Doppler and dynamic scene evolution. For each propagation path, the intersection point with moving objects is updated over time, which makes the path length $L_p(t)$ time-dependent. The relative radial velocity $v_r$ is computed as the projection of the object’s linear velocity onto the ray direction, i.e., the velocity component along the path. This radial velocity induces a Doppler frequency shift $\Delta f = v_r/\lambda$, and the corresponding path contribution is modulated with a term $\exp(j 2\pi \Delta f \, t)$. As a result, each multipath component evolves with time, producing a received signal that naturally contains Doppler spread and time-varying fading.

Dynamic scene simulation further requires updating the underlying acceleration structures. Instead of rebuilding the full BVH at every frame, which scales as $O(N \log N)$ for $N$ primitives, we adopt selective refitting of only the bounding volumes associated with dynamic meshes. Using the MegaGeometry API in OptiX 9, we can efficiently update only the subset of moving objects while reusing the static BVH of the environment. This hybrid strategy reduces overhead significantly and enables simulation of mobility-rich environments at scale.

Through this formulation, the ray tracer outputs not only static coverage maps but also time-evolving channel impulse responses and Doppler power spectra. Moving vehicles, for instance, contribute clear frequency shifts in the channel spectrum, while human motion introduces rapid temporal variations and shadowing effects. These extensions allow the simulator to model realistic wireless channels under mobility while maintaining high computational efficiency.

\vspace{-4pt}
\section{Differentiable Simulation}
In this section, we introduce the detailed implementation of differentiable rendering for the baseline methods: Sionna, Soften Triangle, and our adaptation of path-space differentiable rendering in RF, which also serves as the scattering simulation component of \name.

\vspace{-4pt}

\subsection{Baseline: Soften Triangles}

We implement the soften triangles baseline following Eq.~\eqref{eq:soft_triangle} and Eq.~\eqref{eq:sigmoid_weight}, based on the reparameterized method of images. The core idea is to replace hard visibility with a sigmoid smoothing across different propagation paths. This smoothing introduces geometric approximations that break energy conservation, causing certain paths to either lose or leak energy. A key limitation lies in the choice of the sigmoid parameter $k$: its value is arbitrary and not tied to any physical quantity, but instead depends on the triangle size. As a result, even for the same 3D-scanned scene, different mesh topologies or triangulations may yield inconsistent outcomes. Hence, soften triangles is not a suitable method for implementing differentiable RF rendering.


\vspace{-3pt}
\subsection{Differentiable Signal Processing Algorithms}
\label{appendix:music}
\revise{The differentiability of signal processing algorithms is a crucial prerequisite for achieving fully end-to-end optimization. In this section, we analyze the differentiability of the Multiple Signal Classification (MUSIC) algorithm. Although the MUSIC algorithm is theoretically differentiable, the presence of eigenvalue decomposition leads to ill-conditioned gradients with extremely high variance, rendering it impractical for optimization-based learning tasks.  }

\noindent\textbf{\textit{Forward Processing.}}
The MUSIC algorithm is a high-resolution subspace-based method for estimating the spatial spectrum in MIMO radar systems. The input raw MIMO time-domain complex-valued radar signal $\mathbf{X}(t) \in \mathbb{C}^{M \times N \times T}$ first undergoes range-FFT preprocessing, where
$M$ is the number of transmit antennas, $N$ is the number of receive antennas, 
and $T$ is the number of time samples. After range-FFT, we select the range bin of interest containing the target returns, resulting in a narrowband signal matrix for spatial processing. 
Unlike conventional delay-and-sum beamforming which directly combines the antenna signals with phase shifts, MUSIC leverages the signal subspace structure to achieve super-resolution imaging capability.
The first step involves forming the spatial correlation matrix. We reshape the frequency-domain MIMO signal tensor into a matrix $\mathbf{Y} \in \mathbb{C}^{MN \times T}$ by vectorizing the spatial dimensions. The correlation matrix is then computed as: $\mathbf{R} = \mathbf{Y}\mathbf{Y}^H$,
where $(\cdot)^H$ denotes the conjugate transpose operation. This correlation matrix captures the spatial relationships between different antenna pairs.
The MUSIC algorithm relies on the eigendecomposition of the correlation matrix: $\mathbf{R} = \mathbf{U}\mathbf{\Lambda}\mathbf{U}^H$.
where $\mathbf{U} = [\mathbf{U}_s ; \mathbf{U}_n]$ contains the eigenvectors and $\mathbf{\Lambda}$ contains the eigenvalues in descending order. $\mathbf{U}_s$ spans the signal subspace corresponding to the $K$ largest eigenvalues, while $\mathbf{U}_n$ spans the noise subspace.
The 3D spatial spectrum is computed by evaluating the MUSIC pseudo spectrum at each spatial location $(x,y,z)$:
\begin{equation}
P(x,y,z) = \frac{1}{\mathbf{a}^H(x,y,z)\mathbf{U}_n\mathbf{U}_n^H\mathbf{a}(x,y,z)},
\end{equation}
where $\mathbf{a}(x,y,z)$ is the steering vector that encodes the expected phase relationships for a target at position $(x,y,z)$.

\noindent\textbf{\textit{Differentiation.}}
The MUSIC algorithm presents significant challenges for automatic differentiation due to its reliance on eigendecomposition.

We analyze the differentiability of the \textsc{MUSIC} pseudospectrum. 
Let $\mathbf{R}\in\mathbb{C}^{p\times p}$ be the Hermitian spatial correlation matrix with eigendecomposition
\[
\mathbf{R}=\mathbf{U}\boldsymbol{\Lambda}\mathbf{U}^{H},\qquad
\boldsymbol{\Lambda}=\mathrm{diag}(\lambda_1\ge \cdots \ge \lambda_p),\quad
\mathbf{U}=[\mathbf{u}_1,\ldots,\mathbf{u}_p].
\]
For a chosen model order $K$, we split the basis as $\mathbf{U}=[\mathbf{U}_s,\mathbf{U}_n]$ where $\mathbf{U}_s$ corresponds to the $K$ dominant eigenvalues and $\mathbf{U}_n$ to the remaining ones. The noise projector is $\mathbf{P}_n=\mathbf{U}_n\mathbf{U}_n^H$, and the pseudospectrum is defined by
\begin{equation}
P(\mathbf{a})=\frac{1}{\mathbf{a}^H\mathbf{P}_n\mathbf{a}}
\label{eq:music-spectrum}
\end{equation}
for any steering vector $\mathbf{a}$.

\noindent
\textit{Perturbation of eigenvectors.}
For simple eigenvalues, differentiating the eigenvalue equation $\mathbf{R}\mathbf{u}_i=\lambda_i\mathbf{u}_i$ under the normalization $\mathbf{u}_i^H\mathbf{u}_i=1$ yields the classical first-order perturbation formulas \cite{Kato1995,StewartSun1990}:
\begin{equation}
\mathrm{d}\lambda_i=\mathbf{u}_i^H(\mathrm{d}\mathbf{R})\mathbf{u}_i,
\qquad
\mathrm{d}\mathbf{u}_i=\sum_{j\neq i}\frac{\mathbf{u}_j\,\mathbf{u}_j^H(\mathrm{d}\mathbf{R})\mathbf{u}_i}{\lambda_i-\lambda_j}.
\label{eq:eig-perturb}
\end{equation}
The denominators demonstrate instability when eigenvalue gaps close; at exact degeneracy the derivative is undefined because the eigenspace is invariant only up to a unitary rotation.

\noindent
\textit{Perturbation of projectors.}
It is more stable to differentiate the spectral projector $\mathbf{P}_s=\mathbf{U}_s\mathbf{U}_s^H$ directly. In the eigenbasis, its variation satisfies the Sylvester equation
\begin{equation}
\mathrm{d}\mathbf{P}_s=\mathbf{U}
\begin{bmatrix}
\mathbf{0}&\boldsymbol{\Gamma}\\
\boldsymbol{\Gamma}^H&\mathbf{0}
\end{bmatrix}
\mathbf{U}^H,
\qquad
\boldsymbol{\Lambda}_s\boldsymbol{\Gamma}-\boldsymbol{\Gamma}\boldsymbol{\Lambda}_n
=\mathbf{U}_s^H(\mathrm{d}\mathbf{R})\mathbf{U}_n,
\label{eq:proj-diff}
\end{equation}
so that each entry of $\boldsymbol{\Gamma}$ takes the form
\[
(\boldsymbol{\Gamma})_{ij}=\frac{(\mathbf{U}_s^H(\mathrm{d}\mathbf{R})\mathbf{U}_n)_{ij}}{\lambda_i-\lambda_{K+j}}.
\]
If $\Delta=\lambda_K-\lambda_{K+1}$ denotes the signal-noise gap, then
\begin{equation}
\|\mathrm{d}\mathbf{P}_s\|_2\;\le\;\frac{\|\mathrm{d}\mathbf{R}\|_2}{\Delta},\qquad
\mathrm{d}\mathbf{P}_n=-\mathrm{d}\mathbf{P}_s,
\label{eq:gap-bound}
\end{equation}
which is the infinitesimal form of the Davis-Kahan $\sin\Theta$ theorem \cite{DavisKahan1970}.

\noindent
\textit{Effect on the pseudospectrum.}
Let $s(\mathbf{a})=\mathbf{a}^H\mathbf{P}_n\mathbf{a}$. From \eqref{eq:music-spectrum} and \eqref{eq:gap-bound}, the differential is
\begin{equation}
\mathrm{d}P(\mathbf{a})=-s(\mathbf{a})^{-2}\,\mathbf{a}^H(\mathrm{d}\mathbf{P}_n)\mathbf{a},
\qquad
|\mathrm{d}P(\mathbf{a})|\;\le\;\frac{\|\mathbf{a}\|_2^2}{s(\mathbf{a})^2}\cdot\frac{\|\mathrm{d}\mathbf{R}\|_2}{\Delta}.
\label{eq:music-bound}
\end{equation}
Thus the gradient of the pseudospectrum is well defined whenever $\Delta>0$, but scales like $1/\Delta$ and becomes arbitrarily large near eigenvalue crossings.

\noindent
\textit{Chain to the data matrix.}
Since $\mathbf{R}=\mathbf{Y}\mathbf{Y}^H$, any perturbation obeys
\[
\mathrm{d}\mathbf{R}=(\mathrm{d}\mathbf{Y})\mathbf{Y}^H+\mathbf{Y}(\mathrm{d}\mathbf{Y})^H.
\]
Hence the gradient with respect to the raw signal matrix $\mathbf{Y}$ inherits the same spectral gap dependence. The instability at small $\Delta$ explains the difficulty of incorporating MUSIC into differentiable end-to-end learning systems.

\end{document}